\PassOptionsToPackage{pdfpagelabels=false}{hyperref}
\documentclass[acmsmall,screen,nonacm]{acmart}

\usepackage{graphicx}%
\usepackage{multirow}%
\usepackage{amsmath,amsfonts}%
\usepackage{amsthm}%
\usepackage{xcolor}%
\usepackage{textcomp}%
\usepackage{booktabs}%
\usepackage{listings}%
\usepackage{tikz}

\usetikzlibrary{arrows.meta, quotes, shapes,arrows,matrix,positioning,calc}

\AtBeginDocument{%
  }

\setcopyright{none}

\definecolor{keywordcolor}{rgb}{0.7, 0.1, 0.1}
\definecolor{tacticcolor}{rgb}{0.0, 0.1, 0.6}
\definecolor{commentcolor}{rgb}{0.4, 0.4, 0.4}
\definecolor{symbolcolor}{rgb}{0.0, 0.1, 0.6}
\definecolor{sortcolor}{rgb}{0.1, 0.5, 0.1}
\definecolor{attributecolor}{rgb}{0.7, 0.1, 0.1}

\newcommand{\lean}[1]{\lstinline[language=lean, mathescape=true, basicstyle=\ttfamily\normalsize, breaklines=true]|#1|}

\newcommand{\leanfootnote}[1]{\lstinline[language=lean, mathescape=true, basicstyle=\ttfamily\footnotesize]|#1|}
\newcommand{\leanfootnotealt}[1]{\lstinline[language=lean, mathescape=true, basicstyle=\ttfamily\footnotesize]!#1!}

\begin{document}
\newcommand\ltsnode [3]{\node (#2) [circle, draw, anchor=center, text width={.3cm}, inner sep=1mm, minimum height={.3cm}, align=center, #1] {$#3$};}
\tikzstyle{arrow} = [thick,->,>=stealth,solid, shorten >=2pt, shorten <=2pt,]

\title{A Verified Implementation of the L*-Algorithm in Lean}

\author{Dominik Mascherbauer}
\affiliation{%
  \institution{Johannes Kepler University Linz}
  \city{Linz}
  \country{Austria}}
\email{dominik.mascherbauer@jku.at}

\author{Bernhard K.~Aichernig}
\affiliation{%
  \institution{Johannes Kepler University Linz}
  \city{Linz}
  \country{Austria}}
\email{bernhard.aichernig@jku.at}

\begin{abstract}
Angluin’s L*-algorithm learns the minimal deterministic finite automaton (DFA) of an unknown regular language from membership and equivalence queries to a teacher.
It underpins active automata learning, yet its correctness guarantees have so far rested on pen-and-paper proofs, separate from the implementations in use.
We close this gap for L* using Lean 4, which is both a functional programming language and an interactive theorem prover: we implement L* in the Minimally Adequate Teacher framework and prove its properties directly about the executable code.
We establish two results:
(1)~Termination: for every teacher that answers consistently with a regular language and returns only genuine counterexamples, learning terminates.
(2)~Minimality and correctness: the learned DFA is never larger than the minimal DFA of that language, and if the teacher answers equivalence queries truthfully, it is isomorphic to that minimal DFA.
The result is a single artifact that is both a mechanically checked proof and an executable learner.
To the best of our knowledge, this is the first mechanized correctness proof of the classical L*-algorithm.
\end{abstract}




\maketitle

\section{Introduction}\label{introduction}

Modern software development processes increasingly rely on AI agents to provide code~\cite{chen_evaluating_2021, liventsev_fully_2023}. 
Often, complex algorithms are generated with no formal guarantees about their correctness. 
This leaves one main question: can we trust AI-generated code? 
For critical algorithms, a formal proof provides the strongest form of guarantee and builds trust in an algorithm. 
Historically, algorithm proofs involved a pen-and-paper proof of certain properties, which could then be formalized and checked mechanically in an interactive theorem prover. 
However, executable algorithm implementations, i.e., programs, are mostly not connected to their formal proofs. 
While techniques exist to connect an executable algorithm to a formal proof, they usually involve complex infrastructure for code extraction~\cite{letouzey_new_2003} or external verification of existing code bases~\cite{klein_sel4_2009}, leaving a gap between the proven model and the running artifact. 
If a subtle bug is introduced during implementation or an edge case is missed, it can silently invalidate the theoretical guarantees of the algorithm.

In the field of automata learning, this forms a major challenge, as algorithms like Angluin's L*~\cite{angluin_learning_1987} serve as fundamental building blocks for formal verification and testing~\cite{vaandrager_model_2017}.
For example, learned models were used to reveal security flaws in TLS implementations~\cite{ruiter_protocol_2015}, uncover violations of the respective standards in SSH implementations~\cite{fiterau-brostean_model_2017} and MQTT brokers~\cite{tappler_model-based_2017}, and enable model-based testing of the Git version control system~\cite{muskardin_active_2024}.
Verification and testing, in turn, build trust in complex software systems. 
Consequently, automata learning is a critical step requiring strong formal guarantees. 
While elegant in theory, the L*-algorithm involves managing complex state within its observation table, ensuring properties that allow the construction of a deterministic finite automaton (DFA), while correctly handling responses from a teacher. 
For a correct algorithm implementation, we must rely on getting these details precisely right.

To bridge this gap between algorithm implementation and formal proof, we need a framework that allows both implementing algorithms and proving them correct with an interactive theorem prover. 
Lean 4~\cite{moura_lean_2021} offers such a framework.
It is a functional programming language~\cite{christiansen_functional_2026} with a powerful, integrated interactive theorem prover~\cite{avigad_theorem_2026}. 
This allows us to write the implementation and prove properties about it in the same artifact, entangling the proof and implementation.
Consequently, a bug in the implementation that violates a proven property invalidates the proof, so it cannot go undetected.

In this work, we present an implementation-based formalization of the L*-algorithm in Lean 4. 
We demonstrate how this all-in-one approach closes the gap between proof and implementation. 
A single, trustworthy artifact that is both mechanically checked and executable emerges. 
Our main contributions are:

\begin{enumerate}
    \item A complete, executable implementation of the L*-algorithm in Lean 4, following the Minimally Adequate Teacher (MAT) framework.
    \item A formal proof of the algorithm's correctness: the learned DFA is never larger than the minimal DFA of the target language; for a complete teacher it accepts exactly the target language and is unique up to isomorphism. As in Angluin's paper, the core is the partial-correctness result for closed and consistent observation tables (Theorem~\ref{thm:minimality}).
    \item A formal proof of the algorithm's termination, guaranteeing that the learning process concludes in a finite number of steps for every teacher whose language is regular and whose counterexamples are genuine.
    \item A precise separation of the teacher's guarantees into soundness (genuine counterexamples), which suffices for termination and minimality, and completeness, which is needed for correctness; and a demonstration of the verified implementation with a sound random-word teacher that is fully proven, without any axioms.
\end{enumerate}

The implementation, all proofs, and scripts reproducing the figures stated in this article are available as an artifact~\cite{mascherbauer_aalean_2026}.

The remainder of this article is structured as follows. 
Section~\ref{preliminaries} reviews automata theory, the L*-algorithm, and Lean 4. 
Section~\ref{sec:formalizing-dfas} details our computable data structures. 
In Sections~\ref{sec:teacher} and~\ref{sec:lstar-algo}, we present the implementation of the MAT framework and learning algorithm. 
The formal proofs of correctness and termination are covered in Section~\ref{sec:lstar-proofs}. 
Finally, Section~\ref{sec:run-lstar} demonstrates an example use of our implementation, and we conclude in Section~\ref{sec:conclusion}.

\section{Preliminaries}\label{preliminaries}
This section introduces the basic terminology on automata theory foundational to this work. 
Additionally, we discuss active automata learning and the Lean 4 programming language, which are the main topics this work is based on.

\subsection{Finite State Machines}\label{sec:fsm}
A Finite State Machine (FSM) represents a system with a finite number of states and transitions between those states. 
FSMs are used in various domains, such as digital circuit design~\cite{minns_fsm-based_2008, harris_digital_2021} and network protocols~\cite{pferscher_mining_2023, tappler_model-based_2017}.

In this article, we are mainly interested in deterministic finite automata (DFAs), which are used to recognize regular languages. 
Formally, a DFA is represented by a 5-tuple~\cite{sipser_introduction_2013}:
\begin{itemize}
\item $Q$: a finite set of states
\item $A$: a finite set of symbols called the alphabet
\item $\delta: Q\times A \rightarrow Q$: the transition function
\item $q_0 \in Q$: the initial state
\item $F \subseteq Q$: a finite set of accepting states
\end{itemize}

As $\delta$ is a total function, a DFA is always \emph{complete}, meaning it has a defined transition for every symbol in its alphabet from every state. 
A \emph{word} is a sequence of symbols, also known as a \emph{string}. 
It is \emph{accepted} by a given DFA if the word transitions from the initial state $q_0$ through the DFA and ends in an accepting state in $F$. 
The set of all words accepted by a DFA is called its \emph{language}. 
Languages that can be represented by a DFA are called \emph{regular languages} and form the lowest level in the Chomsky hierarchy.

\subsection{Active Automata Learning}\label{sec:active-learning}
Active automata learning describes the process of inferring an unknown automaton from a teacher. 
Angluin~\cite{angluin_learning_1987} introduced an active automata learning framework, formalized in the Minimally Adequate Teacher (MAT) framework, and introduced the L*-algorithm as a prime example. 
The goal of the L*-algorithm is to learn an unknown regular language and produce its minimal DFA. 
L* was later extended to work with different automata types, such as Mealy machines~\cite{shahbaz_inferring_2009} and Moore machines, and even Markov Decision Processes (MDPs)~\cite{tappler_l-based_2021}. 
However, as this article covers a mechanical proof of Angluin's original L*-algorithm, this work focuses on learning a DFA representation.

\subsubsection{Minimally Adequate Teacher Framework}\label{sec:mat}
\begin{figure}
    \centering
	\begin{tikzpicture}[node distance=6cm, auto]
		\node (teacher) [rectangle, draw, minimum width=2cm, minimum height=4cm] {Teacher};
		\node (learner) [rectangle, draw, minimum width=2cm, minimum height=4cm, right=of teacher] {Learner};
					
		\draw[arrow] ([yshift=1.5cm]learner.west) -- ([yshift=1.5cm]teacher.east) node[midway, above] {Membership Query (word)};
		\draw[arrow] ([yshift=1cm]teacher.east) -- ([yshift=1cm]learner.west) node[midway, below] {yes / no};
					
		\draw[arrow] ([yshift=-1cm]learner.west) -- ([yshift=-1cm]teacher.east) node[midway, above] {Equivalence Query (hypothesis)};
		\draw[arrow] ([yshift=-1.5cm]teacher.east) -- ([yshift=-1.5cm]learner.west) node[midway, below] {yes / counterexample};
	\end{tikzpicture}
    \caption{Minimally Adequate Teacher Framework}
    \label{fig:MAT-framework}
\end{figure}
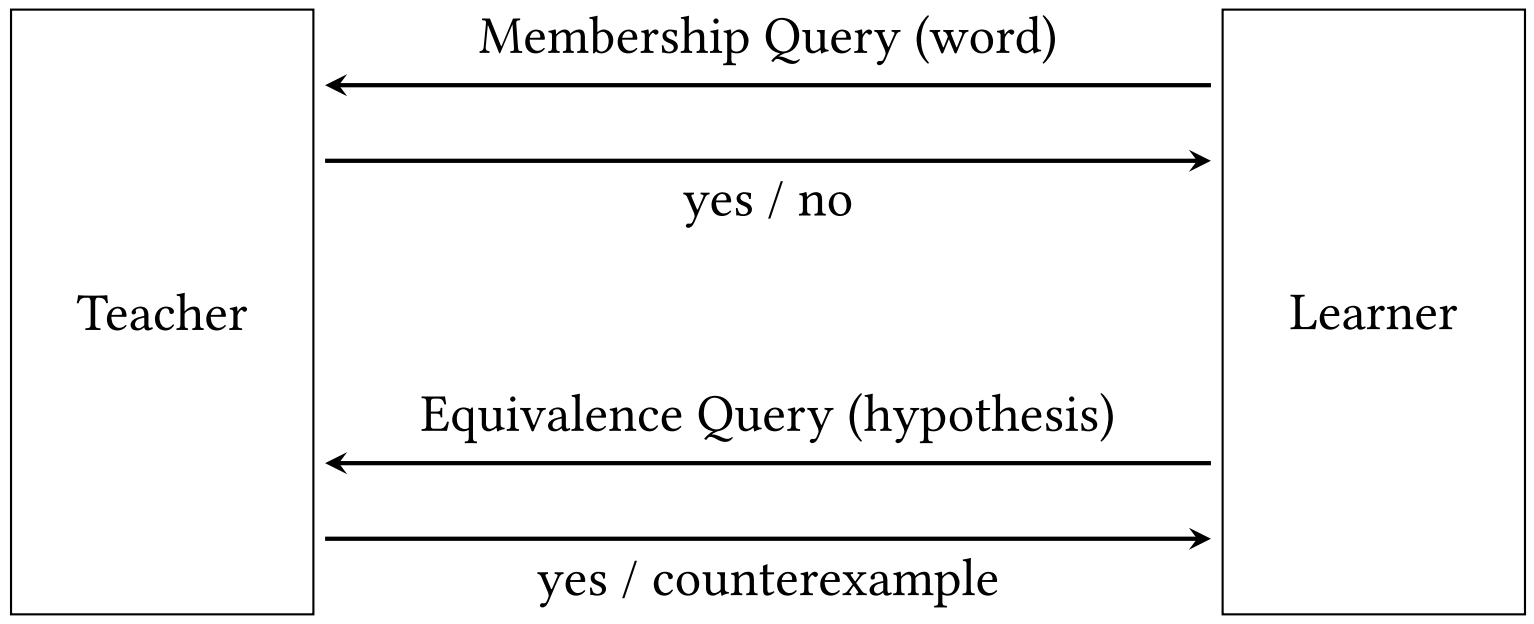

A central part of the L*-algorithm is the MAT framework, shown in Figure~\ref{fig:MAT-framework}. 
It defines the queries a learner can ask a teacher:
\begin{itemize}
    \item \textbf{Membership Query:} 
    The learner asks whether a word $w$ is in the unknown regular language. 
    This is answered with a simple boolean decision, yes or no.
    \item \textbf{Equivalence Query:} 
    The learner produces a hypothesis from its observations so far and presents it to the teacher. 
    The teacher then responds with either a counterexample if the hypothesis does not match the unknown regular language, or confirms its correctness if it does.
\end{itemize}

In our formalization, we separate these guarantees.
A teacher only has to return genuine counterexamples, and completeness (confirming only correct hypotheses) is an additional assumption (Section~\ref{sec:teacher}).

\subsubsection{Observations}\label{sec:prelim-observations}
During the learning process, the L* learner makes observations on the unknown regular language. 
These observations are stored in a data structure called an \emph{observation table}. 
An observation table consists of a prefix-closed set $S$ of strings used to identify states, and a suffix-closed string set $E$ of experiments used to distinguish them. 
A prefix-closed set is a set that contains all prefixes of all its words, while a suffix-closed set is a set that contains all suffixes of all its words.

The table also considers the set of one-symbol extensions of $S$, denoted $S \cdot A$. 
A row for a string $s$, denoted $row(s)$, is a function mapping each experiment $e \in E$ to the result of the membership query for the concatenated string $s \cdot e$. 
The function $T$ stores these results, such that $T(s \cdot e)$ is $1$ if $s \cdot e$ is in the unknown language, and $0$ otherwise.

Table~\ref{tab:obs-table-prelim} depicts an example observation table with alphabet $\{a,b\}$ that could appear during a learning process. 
The set $S$ contains $\{\epsilon, a\}$ and the set $E$ contains $\{\epsilon, a\}$. 
The table also shows the rows for the extended set, $S \cdot A$. 
To get a specific cell of this table, we can use the function $T$ mapping the concatenated strings of $S \cup S\cdot A$ and $E$ to their Boolean cell values, e.g., $T(a\cdot \epsilon) = T(b\cdot \epsilon) = 0$.
A row maps to all observations for a string in $S \cup S \cdot A$, e.g., $row(a) = row(b) = [0,1]$.

\begin{figure}
\begin{minipage}[c]{0.48\linewidth}
    \centering
    \begin{tabular}{cc|cc}
        \multicolumn{2}{c|}{\multirow{2}{*}{T}}                       & \multicolumn{2}{c}{E} \\ \cline{3-4} 
        \multicolumn{2}{c|}{}                                         & $\epsilon$ & a        \\ \hline
        \multicolumn{1}{c|}{\multirow{2}{*}{S}}          & $\epsilon$ & 1          & 0        \\
        \multicolumn{1}{c|}{}                            & a          & 0          & 1        \\ \hline
        \multicolumn{1}{c|}{\multirow{3}{*}{S$\cdot$A}}  & b          & 0          & 1        \\
        \multicolumn{1}{c|}{}                            & aa         & 1          & 0        \\
        \multicolumn{1}{c|}{}                            & ab         & 0          & 1
    \end{tabular}
    \captionof{table}{A closed and consistent Observation Table.}
    \label{tab:obs-table-prelim}
\end{minipage}\hfill
\begin{minipage}[c]{0.48\linewidth}
    \centering
    \begin{tikzpicture}[node distance=2cm]
        \ltsnode{}{q0}{q0}
        \ltsnode{right of = q0}{q1}{q1}
        
        \node [circle, minimum height=.25cm,text width = .25cm, anchor=center,draw] at (q0.center){};
        
        \draw [arrow] ([yshift=2pt]q0.east) -- ([yshift=2pt]q1.west) node [midway, above] {\footnotesize$\mathsf{a,b}$};
        \draw [arrow] ([yshift=-2pt]q1.west) -- ([yshift=-2pt]q0.east) node [midway, below] {\footnotesize$\mathsf{a}$};
        \path [arrow] (q1) edge[loop above] node[above=1pt] {\footnotesize$\mathsf{b}$} (q1);
        
        \draw [arrow] node[auto] {} ++ (0,1) to (q0);
    \end{tikzpicture}
    \caption{The DFA created from Table~\ref{tab:obs-table-prelim}.}
    \label{fig:obs-table-prelim-dfa}
\end{minipage}
\end{figure}

An observation table has two important properties, closedness and consistency:
\begin{itemize}
    \item It is closed if for every string in $S \cdot A$, its row is already present for some string in $S$.
    \item It is consistent if for any two strings $s_1, s_2 \in S$ with $row(s_1) = row(s_2)$, their one-symbol extensions also have identical rows, i.e., for all $a \in A$, $row(s_1\cdot a) = row(s_2\cdot a)$.
\end{itemize}

To create a hypothesis DFA, an observation table must be closed and consistent. 
The states of the DFA correspond to the unique rows for strings in $S$. 
The transition from a state $row(s)$ on symbol $a$ is the state corresponding to $row(s\cdot a)$. 
Closedness ensures that all known states are captured in the states of $S$ and that all transitions connect two known states. 
Consistency ensures that the transitions are well-defined.  
A state is accepting if its row has an output of 1 for the empty experiment $\epsilon$. 

The observation table in Table~\ref{tab:obs-table-prelim} is already closed and consistent. 
The DFA in Figure~\ref{fig:obs-table-prelim-dfa} shows the corresponding automaton. 
Each unique row from the $S$ set becomes a state. 
$row(\epsilon) = [1,0]$ becomes state $q_0$, and $row(a) = [0,1]$ becomes state $q_1$. 
Since $T(\epsilon \cdot \epsilon) = 1$, state $q_0$ is accepting. 
To get the transitions from a state, we check its one-symbol extensions, e.g., for $q_1$, $row(a\cdot a) = [1,0]$ corresponds to $q_0$, and $row(a\cdot b) = [0,1]$ corresponds to $q_1$.

\subsection{Myhill-Nerode Theorem}\label{sec:myhill-nerode}
The theoretical foundation of the L*-algorithm is the Myhill-Nerode theorem~\cite{sipser_introduction_2013}. 
This theorem provides a characterization of regular languages and guarantees the existence of a minimal DFA, unique up to isomorphism, for any regular language.

The theorem is based on the Nerode congruence, an equivalence relation $\equiv_L$ defined for a language $L$. 
Two strings $u$ and $v$ are considered indistinguishable with respect to $L$, written $u \equiv_L v$, if for any suffix $z$, appending it to both $u$ and $v$ results in strings that are either both in $L$ or both not in $L$. 
If a suffix $z$ exists that separates them, e.g., $u \cdot z \in L$ but $v \cdot z \notin L$, it is called a distinguishing suffix.
\[ u \equiv_L v \iff \forall z \in A^*, (u \cdot z \in L \iff v \cdot z \in L) \]
The Myhill-Nerode theorem states that a language $L$ is regular if and only if the Nerode congruence $\equiv_L$ has a finite number of equivalence classes (a finite index).
Furthermore, the number of states in the minimal DFA for $L$ is exactly equal to the number of these equivalence classes.

The L*-algorithm can be seen as an algorithm that systematically discovers these equivalence classes. 
The set $S$ contains representative strings for each discovered class, and the set $E$ contains the distinguishing suffixes $z$ (experiments) needed to separate them. 
The rows in a closed, consistent observation table correspond directly to the algorithm's current approximation of the Nerode equivalence classes.

\subsection{Lean 4}\label{sec:lean}
Lean 4~\cite{moura_lean_2021} is a functional programming language~\cite{christiansen_functional_2026} that also serves as an interactive theorem prover~\cite{avigad_theorem_2026}. 
Combining both worlds, we can prove source code implemented in Lean 4 correct within the same source file, using Lean's theorem prover to reason about the code's behavior.

\subsubsection{Dependent Type Theory}\label{sec-dependent-type-theory}
Lean is based on a formal logical foundation known as dependent type theory. 
A key idea is that types can depend on values. 
For example, one can define a type $\mathit{Vector}\ \alpha\ n$ representing lists of type $\alpha$ of length $n$, where the type itself depends on the value $n$. 
This allows properties to be encoded directly in the types of functions.
For instance, a function concatenating a $\mathit{Vector}\ \alpha\ n$ and a $\mathit{Vector}\ \alpha\ m$ would be typed to return a $\mathit{Vector}\ \alpha\ (n+m)$, proving the correct output length at compile time. 

Another core concept is the Curry-Howard correspondence, where propositions are types and proofs are terms of that type~\cite{wadler_propositions_2015}. 
A proof of $b \neq 0$ is a value whose type encodes this proposition.
Proofs are erased during compilation, so data and proofs can be mixed freely. 
However, not every definition is computable.
Definitions that rely on classical reasoning, such as \lean{Teacher.maxStates} (Section~\ref{sec:teacher}), are marked \lean{noncomputable} and cannot be executed. 
Consequently, when implementing an algorithm, one must be careful to use only computable constructs to ensure the final implementation is executable.

\subsubsection{Lean 4 Type Classes}\label{sec:type-class}
With type classes, Lean provides a way to use ad-hoc polymorphism. 
When a function requires a type to have certain capabilities, defined by a type class, Lean can automatically find a corresponding implementation (an instance) from the context. 
This allows certain information about types and variables to be inferred without the need to pass them directly as arguments.

In addition to type classes, Lean also supports polymorphic structure types with inheritance. A structure is a user-defined type that groups several named fields together.
Inheritance allows adding additional values and properties similar to a type class. 
However, 
in contrast to type classes,
an instance of a structure cannot be implicitly synthesized from the context. Their instances need to be passed explicitly as arguments to a function.

To aid readability and usability of the final L*-algorithm implementation, we introduce a type class \lean{Alphabet} for symbol types (Section~\ref{sec:alphabet-repr}) and structures such as \lean{LearnedDFA}, \lean{ObservationTable} and \lean{Teacher} that bundle data with the properties they guarantee.

\subsubsection{Implicit Parameters}\label{sec:implicit-param}
Another extensively used concept in this article is implicit parameters. 
From a high-level point of view, this allows the inference of certain parameters from the context or other parameters, increasing convenience when calling functions.

Lean 4 supports multiple different variants of implicit parameters~\cite{noauthor_implicit_2026}, two of which were used in our implementation:
\begin{itemize}
    \item \textbf{Ordinary implicit parameters:} 
    Function parameters that should be determined via unification. 
    A call site should have one potential argument value for the implicit parameter to make the function call well-typed. 
    The Lean elaborator tries to find all implicit arguments of the function. 
    In the function signature, an implicit parameter is written in curly braces, e.g., \lean{def dummy \{a: Nat\}: ...}.

    \item \textbf{Instance implicit parameters:} 
    Found via type class synthesis and written in square brackets, e.g., \lean{def dummy (a: Type) [BEq a]: ...}. 
    Usually, this kind of parameter is written without an explicit name and specifies a type class instance for a type parameter.
    We name it when the instance is used directly, e.g., \lean{[alphabet: Alphabet Symbol]} with \lean{alphabet.symbols} (see Section~\ref{sec:learned-dfa-structure}).
    These implicit parameters are provided via the local context or by Lean's type class synthesis mechanism.
\end{itemize}

Take the following signature for a safe division for example. 
It defines a function that requires two natural numbers, \lean{a} and \lean{b}, and a proof that the second number is unequal to \lean{0}, \lean{h}.

\begin{lstlisting}
def safeDiv (a b: Nat) (h: b ≠ 0): Nat := a / b
\end{lstlisting}

Using implicit parameters, we can rewrite this as follows.

\begin{lstlisting}
def safeDiv (a: Nat) {b: Nat} (h: b ≠ 0): Nat := a / b
\end{lstlisting}

Now we have an implicit parameter for \lean{b}, meaning the argument value of \lean{b} will be inferred from the parameter \lean{h}.
We can exploit the fact that dividing by a successor of a natural number is always safe using the theorem \lean{succ_ne_zero'}, which states that any successor of a natural number is unequal to \lean{0}, and which we prove with \lean{simp}.
In Lean, the keyword \lean{by} activates the \emph{proof tactic mode}, which allows closing a goal sequentially by so-called \emph{tactics}.
One such \emph{tactic} is \lean{simp}, which tells Lean to simplify the goal state as much as possible applying theorems marked for use with \lean{simp}.

For example, if we call this function with \lean{10} and \lean{succ_ne_zero' 1}, a proof of \lean{Nat.succ 1 ≠ 0}, Lean infers the parameter \lean{Nat.succ 1} for \lean{b} resulting in the division \lean{10/2}.

\begin{lstlisting}
theorem succ_ne_zero' (n: Nat): n.succ ≠ 0 := by simp
def result := safeDiv 10 (succ_ne_zero' 1)  -- Divide 10 by the successor of 1
\end{lstlisting}

When computing \lean{result} in the listing above, the term elaborator first unifies the function call to its canonical form\footnote{Prefixing a function with \leanfootnote{@} makes all its implicit parameters explicit, so they can be passed as ordinary arguments.}:
\begin{lstlisting}
@safeDiv 10 (Nat.succ 1) (succ_ne_zero' 1): Nat
\end{lstlisting}
The canonical form is checked by the Lean type checker and executed by Lean's runtime to provide the value \lean{result: Nat = 5}.

\subsubsection{Mathlib}\label{sec:mathlib}
Mathlib~\cite{the_mathlib_community_lean_2020} is a comprehensive Lean library providing proofs for many fields of mathematics, including set theory, algebra, and analysis.
In particular, we use mathlib's \lean{DFA} structure, regular languages (\lean{Language.IsRegular}), and its formalization of the Myhill-Nerode theorem in terms of left quotients (\lean{Language.leftQuotient}), which bounds the number of states in our termination and minimality proofs.
Using mathlib as a foundation saved us from having to formalize these standard definitions from scratch and allowed us to focus on the L*-algorithm itself. 
This allows us to reuse many well-documented and actively maintained theorems and structures, which immensely simplifies writing proofs for the algorithm's correctness and termination.

\section{Formalizing DFAs: Reusability vs. Computability}\label{sec:formalizing-dfas}
Many structures in libraries like mathlib are specialized for proofs and stay close to their theoretical definitions. 
For our goal of proving an executable implementation correct, such structures present a major disadvantage: many operations on them are not computable and thus cannot be compiled into executable code. 
In mathlib, for example, a \lean{Set} is defined as a predicate \lean{$ $α → Prop}, making operations like iteration impossible to compute.
To circumvent this, we use computable representations, such as \lean{List} instead of mathlib's non-computable \lean{Set}, for our algorithm's data structures.

In the L*-algorithm implementation, we decided to reuse as many available structures from mathlib as possible. 
This aligns with the principle of reusability, especially with regards to the broader Lean ecosystem including libraries like mathlib and CSLib~\cite{barrett_cslib_2026}. 
Furthermore, this allows us to leverage many theorems, which immensely simplified the correctness proofs for our algorithm.

\subsection{Alphabet Representation}\label{sec:alphabet-repr}

The first example where we rely on lists rather than sets is the alphabet of a DFA. The reason is that for certain parts of the L*-algorithm, such as exploring a constructed DFA, we need to iterate over all symbols in the alphabet. 
It is formalized as a type class \lean{Alphabet}.
In mathlib, a \lean{DFA}'s alphabet is a polymorphic type \lean{Symbol: Type}. 
Our \lean{Alphabet} type class is parameterized by this same \lean{Symbol} type, allowing us to augment the \lean{DFA} with a computable representation\footnote{Listings of AALean declarations are taken from the artifact~\cite{mascherbauer_aalean_2026}; proofs are elided with \leanfootnote{...}, and listings that are simplified are marked as such in the text.}:

\begin{lstlisting}
class Alphabet (Symbol: Type) where
  symbols: List Symbol
  prop: (∀ symbol: Symbol, symbol ∈ symbols) ∧ symbols.Nodup ∧ ¬symbols.isEmpty
  toDecidableEq: DecidableEq Symbol
\end{lstlisting}

The type class \lean{Alphabet} uses the following values and properties:
\begin{itemize}
    \item \lean{Symbol}: The polymorphic type representing the alphabet.
    \item \lean{symbols}: The list of all symbols in \lean{Symbol}.
    \item \lean{prop}: The property of the list representation connecting the list representation to the underlying alphabet type,
    comprising (1) each element in \lean{Symbol} is in the list representation, (2) the list representation does not have any duplicates, and (3) the list is not empty. Properties (1) and (2) allow using the list representation, and therefore the alphabet type, as a finite set in proofs; (3) is needed to draw random symbols (Section~\ref{sec:random-word-teacher}).
    \item \lean{toDecidableEq}: A \lean{DecidableEq} instance for the alphabet type to provide computable equality comparisons between symbols.
\end{itemize}

To construct a usable alphabet for the L*-algorithm implementation, we must create an instance of \lean{Alphabet} for the concrete alphabet type. 
The Listing below shows how to create such an instance for a simple inductive type. 
We provide the list of all symbols and then prove that this list contains every element of the type exactly once and is not empty.

\noindent\begin{minipage}{\linewidth}
\begin{lstlisting}
inductive Syms where
| a | b
deriving DecidableEq

instance: Alphabet Syms where
  symbols := [.a,.b]
  prop := by ...
  toDecidableEq := inferInstance
\end{lstlisting}
\end{minipage}
The Lean command \lean{deriving DecidableEq} tells Lean to automatically derive a decision procedure for the equality of the inductive type \lean{Syms}. 
For \lean{symbols}, \lean{.a} and \lean{.b} are the short form of \lean{Syms.a} and \lean{Syms.b}.
For brevity we omit the proof of \lean{prop}.
Since \lean{Syms} derives \lean{DecidableEq}, Lean can automatically infer the required \lean{DecidableEq Syms} instance with \lean{inferInstance}.

\subsection{Representing DFAs Learned by the L*-Algorithm}\label{sec:learned-dfa}
The mathlib library provides a standard definition for a DFA, shown in the Listing below. 
It defines the structure \lean{DFA} via a state type (\lean{$ $σ}), a transition function (\lean{step}), a starting state (\lean{start}), and a set of accepting states (\lean{accept}).
Additionally, mathlib also defines the function \lean{DFA.eval: List α → σ}, evaluating a DFA with a sequence of inputs starting at the initial state \lean{start} step-by-step using the function \lean{step}.

\begin{lstlisting}
structure DFA (α : Type u) (σ : Type v) where
  step : σ → α → σ  -- transition function
  start : σ            -- initial state
  accept : Set σ       -- set of accepting states
\end{lstlisting}

While this structure is ideal for proofs, its reliance on \lean{Set σ} for the accepting states and the polymorphic type \lean{$ $σ} for the state space makes it unsuitable for a fully computable implementation. 
The \lean{accept} field is a predicate \lean{$ $σ → Prop}, which cannot be iterated over, and there is no built-in way to enumerate all states in \lean{$ $σ}. 
This prevents Lean from compiling parts of an implementation that rely on these operations, such as creating a string representation of a DFA for display.

Furthermore, our implementation of L* produces computable lists of states and accepting states from an observation table which is not directly reflected by the mathlib \lean{DFA} structure. To bridge this gap, we introduce the \lean{LearnedDFA} structure, which serves as a computable representation of the \lean{DFA} structure.

\subsubsection{LearnedDFA Structure Definition}\label{sec:learned-dfa-structure}

As shown in the Listing below, it bundles the core components of \lean{DFA} with explicit, computable \lean{List} representations for its states and accepting states.

\begin{lstlisting}
structure LearnedDFA (Symbol State: Type) [alphabet: Alphabet Symbol] where
  states: List State
  accept: List State
  start: State
  step: State → Symbol → State

  toDecidableEqState: DecidableEq State

  prop_states: ¬states.isEmpty ∧ states.Nodup ∧
    ∀ state ∈ states, ∃ w: List Symbol, w.foldl (fun state' a => step state' a) start = state
  prop_accept: ∀ state ∈ accept, state ∈ states
  prop_start: start ∈ states
  prop_step: ∀ state ∈ states, ∀ symbol ∈ alphabet.symbols, step state symbol ∈ states
\end{lstlisting}

A \lean{LearnedDFA} contains the following fields and properties needed to define a DFA and map it to its mathlib counterpart:
\begin{itemize}
    \item \lean{states}: The list representation of the states of a \lean{LearnedDFA}.
    \item \lean{accept}: The list of accepting states in the \lean{LearnedDFA}.
    \item \lean{start}: The initial state of the \lean{LearnedDFA}.
    \item \lean{step}: The transition function of the \lean{LearnedDFA}.
    \item \lean{toDecidableEqState}: A \lean{DecidableEq} instance for the \lean{LearnedDFA}'s state type. 
    This allows deciding equality of states of a \lean{LearnedDFA} as a boolean value.
    \item The four properties, \lean{prop_states}, \lean{prop_accept}, \lean{prop_start}, and \lean{prop_step}, ensure well-formedness of the \lean{LearnedDFA}, i.e., the list of states is non-empty, contains no duplicates, every state is reachable from the start, start and accepting states must be elements of the state list, and the transition function must be closed over the list of states.
\end{itemize}

The \lean{LearnedDFA}'s alphabet is the instance-implicit argument \lean{[alphabet: Alphabet Symbol]} of the structure (Section~\ref{sec:implicit-param}), which \lean{prop_step} uses.

\subsubsection{Connection to Mathlib DFAs}\label{sec:learned-dfa-mathlib}
Each learning round within the L*-algorithm implementation produces a new \lean{LearnedDFA}. 
To leverage mathlib for proofs, we provide two mapping functions from our \lean{LearnedDFA} to a mathlib \lean{DFA}, as shown in the Listing below:

\begin{itemize}
    \item \lean{Coe (LearnedDFA Symbol State) (DFA Symbol State)}: An instance of Lean's \lean{Coe} type class. 
    This allows for coercions from a \lean{LearnedDFA} to a mathlib \lean{DFA}. 
    In Lean, this coercion instance is automatically used when some variable has type \lean{LearnedDFA} but it is used in a context where \lean{DFA} is expected. 
    Coercion allows implicitly calling the function \lean{coe} specified within the instance if needed.

    \item \lean{LearnedDFA.toAttachedDFA}: 
    The \lean{Coe} instance supports automated type coercion but is in some cases too simple: it ignores the \lean{states} list of the \lean{LearnedDFA}. This is possible, since mathlib's \lean{DFA} encodes the state space as a general type parameter ($\sigma$). Note that this would even allow constructing a DFA with an infinite state space, e.g., with \lean{Nat} as the state type.

    With the function \lean{LearnedDFA.toAttachedDFA}, we also map the states from the \lean{LearnedDFA} to the mathlib \lean{DFA}.
    We do this by attaching the state-membership information to the \lean{DFA}'s state type using a Lean subtype \lean{\{state // state ∈ ldfa.states\}}.
    This type requires that each state value comes with a proof that it belongs to the list of learned states.
    The value and property are accessible with \lean{.val} and \lean{.prop}, respectively.
    We can see in the listing that each constructed state is now a pair, denoted with angle brackets, of the state value and its membership proof.
\end{itemize}

\begin{lstlisting}
instance: Coe (LearnedDFA Symbol State) (DFA Symbol State) where
  coe ldfa := {
    step := fun state symbol => ldfa.step state symbol
    start := ldfa.start
    accept := {state | state ∈ ldfa.accept}
  }

def LearnedDFA.toAttachedDFA (ldfa: LearnedDFA Symbol State): DFA Symbol {state // state ∈ ldfa.states} := {
    step := fun state symbol => ⟨ldfa.step state.val symbol, ldfa.prop_step state.val state.prop symbol (alphabet.prop.left symbol)⟩
    start := ⟨ldfa.start, ldfa.prop_start⟩
    accept := {⟨state.val, ldfa.prop_accept state.val state.prop⟩ | state ∈ ldfa.accept.attach}
  }
\end{lstlisting}

When traversing a DFA, e.g., for proving that two sequences end in the same state, most of the time the coercion \lean{DFA} type without the additional property information is enough and reduces complexity of proofs significantly.
However, in certain instances, such as providing a bijective function between two mathlib \lean{DFA}s with the same number of states, we need the additional information of a \lean{DFA}'s state space to create a one-to-one mapping between states.

\subsubsection{Instantiation and Use of LearnedDFAs}\label{sec:learned-dfa-inst}
\begin{figure}
    \centering
    \begin{tikzpicture}[node distance=2cm]
        \ltsnode{}{q0}{\text{q0}}
        \ltsnode{right of = q0}{q1}{\text{q1}}
        \ltsnode{below of = q0}{q2}{\text{q2}}
        \ltsnode{right of = q2}{q3}{\text{q3}}
        
        \node [circle, minimum height=.25cm,text width = .25cm, anchor=center,draw] at (q0.center){};
        
        \draw [arrow] ([yshift=2pt]q0.east) -- ([yshift=2pt]q1.west) node [midway, above] {\footnotesize$\mathsf{a}$};
        \draw [arrow] ([xshift=-2pt]q0.south) -- ([xshift=-2pt]q2.north) node [midway, left] {\footnotesize$\mathsf{b}$};
        \draw [arrow] ([yshift=-2pt]q1.west) -- ([yshift=-2pt]q0.east) node [midway, below] {\footnotesize$\mathsf{a}$};
        \draw [arrow] ([xshift=-2pt]q1.south) -- ([xshift=-2pt]q3.north) node [midway, left] {\footnotesize$\mathsf{b}$};
        \draw [arrow] ([yshift=2pt]q2.east) -- ([yshift=2pt]q3.west) node [midway, above] {\footnotesize$\mathsf{a}$};
        \draw [arrow] ([xshift=2pt]q2.north) -- ([xshift=2pt]q0.south) node [midway, right] {\footnotesize$\mathsf{b}$};
        \draw [arrow] ([yshift=-2pt]q3.west) -- ([yshift=-2pt]q2.east) node [midway, below] {\footnotesize$\mathsf{a}$};
        \draw [arrow] ([xshift=2pt]q3.north) -- ([xshift=2pt]q1.south) node [midway, right] {\footnotesize$\mathsf{b}$};
        
        \draw [arrow] node[auto] {} ++ (-1,0) to (q0);
    \end{tikzpicture}
    \caption{Visual representation of a \lean{LearnedDFA} instance, recognizing words with an even number of \lean{a}s and an even number of \lean{b}s.}
    \label{fig:learned-dfa}
\end{figure}
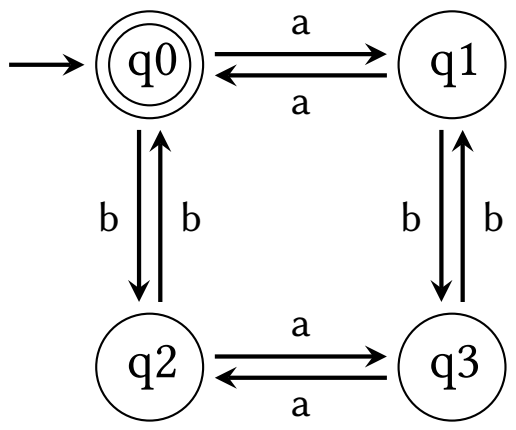

To instantiate a \lean{LearnedDFA} object, we must provide its components (states, accepting states, initial state, and transition function), along with proofs of its well-formedness properties. 
The Listing below shows an example using the \lean{Syms} alphabet and \lean{String} as the state type, with the resulting automaton visualized in Figure~\ref{fig:learned-dfa}: 

\noindent\begin{minipage}{\linewidth}
\begin{lstlisting}
def ldfa: LearnedDFA Syms String where
  states := ["q0", "q1", "q2", "q3"]
  accept := ["q0"]
  start := "q0"
  step state symbol := match state, symbol with (*@\label{ln:ldfa-step}@*)
    | "q0", .a => "q1"
    | "q0", .b => "q2"
    | "q1", .a => "q0"
    | "q1", .b => "q3"
    | "q2", .a => "q3"
    | "q2", .b => "q0"
    | "q3", .a => "q2"
    | "q3", .b => "q1"
    | _, _ => "∅"  -- unreachable from the states above (see `prop_step`)
  toDecidableEqState := inferInstance
  prop_states := by ...
  prop_accept := by simp
  prop_start := by simp
  prop_step := by simp [Alphabet.symbols]
\end{lstlisting}
\end{minipage}

The \lean{step} function in Line~\ref{ln:ldfa-step} is implemented using pattern matching. 
Since pattern matching must be exhaustive, we include a catch-all case at the end. 
However, because \lean{prop_step} guarantees that \lean{step} is only ever called with states from \lean{states}, this catch-all case is unreachable in valid executions.

Except for \lean{prop_states}, which needs a word reaching each state (omitted here), the proofs for the properties are straightforward for this simple \lean{LearnedDFA} and can be discharged with the \lean{simp} tactic.
For \lean{prop_step}, we instruct \lean{simp} to unfold \lean{symbols} from the \lean{Alphabet} instance to ensure it can check all transitions exhaustively.

\section{The Minimally Adequate Teacher}\label{sec:teacher}

The L*-algorithm operates within the Minimally Adequate Teacher (MAT) framework, introduced by Dana Angluin~\cite{angluin_learning_1987}. 
In this framework, a \textit{teacher} that knows a regular language $L$ answers two types of queries from a \textit{learner}.
A membership query asks whether a given word is in $L$. 
An equivalence query takes a hypothesis DFA from the learner.
If the hypothesis is correct, the teacher confirms it, otherwise, it returns a counterexample, a word that distinguishes the hypothesis from the true language.

\subsection{The Teacher Structure}\label{sec:teacher-structure}

In our Lean implementation, we capture the teacher's behavior in the MAT framework with the \lean{Teacher} structure, shown in the Listing below:

\begin{lstlisting}
structure Teacher (Symbol: Type) [Alphabet Symbol] where
  membershipQuery: List Symbol → Bool
  equivalenceQuery: {State: Type} → LearnedDFA Symbol State → Option (List Symbol)

  prop_L: Language.IsRegular {w: List Symbol | membershipQuery w}
  prop_cex: ∀ {State: Type} (ldfa: LearnedDFA Symbol State) (cex: List Symbol),
    equivalenceQuery ldfa = some cex → ¬(cex ∈ ldfa.accepts ↔ membershipQuery cex)
\end{lstlisting}

A \lean{Teacher} is parameterized by a \lean{Symbol} type for the alphabet.
The \lean{equivalenceQuery} accepts a hypothesis over any state type (\lean{\{State: Type\}}), such that one teacher works for all hypotheses.
The \lean{[Alphabet Symbol]} instance ensures we have a computable list representation of the alphabet. 
The structure contains both computable functions for the learner and proof-oriented properties that guarantee its correctness, as follows:

\subsubsection{Computable Queries}
The learner interacts with a \lean{Teacher} exclusively through two functions:

\begin{itemize}
    \item \lean{membershipQuery}: This function takes a word (as a \lean{List Symbol}) and returns {\it true} if the word is in the language \lean{L}, and {\it false} otherwise. 
    The learner uses this to fill its observation table.
    \item \lean{equivalenceQuery}: This function takes a \lean{LearnedDFA} as a hypothesis. 
    We use our computable \lean{LearnedDFA} structure here, as opposed to a mathlib \lean{DFA}, because a concrete teacher implementation may need to iterate over the hypothesis's states to find a counterexample, an operation that requires a computable data structure.
    The query returns \lean{none} if it finds no counterexample, or \lean{some cex} containing a counterexample word.
\end{itemize}

\subsubsection{Proof-Oriented Properties and Definitions}
The following fields are not used by the learner's executable implementation but are essential for proving its correctness:

\begin{itemize}
    \item \lean{prop_L}: The language of \lean{membershipQuery} is regular. 
    This guarantees that a minimal DFA for the language exists (Section~\ref{sec:myhill-nerode}).
    \item \lean{prop_cex}: Every returned counterexample is genuine (hypothesis and membership query disagree on it).
\end{itemize}

The Listing below shows two further definitions about teachers.
\lean{Teacher.L} defines the abstract, regular language known by the teacher as the set of all words accepted by the \lean{membershipQuery}.
\lean{Teacher.Complete n} states that the teacher is complete for hypotheses with at most \lean{n} states if returning \lean{none} for such a hypothesis implies that it accepts exactly \lean{teacher.L}. 
This is not a field, since a teacher that tests random words cannot guarantee it. 
Completeness is only assumed where it is needed, for the correctness of the result (Section~\ref{sec:lstar-proofs}).

\noindent\begin{minipage}{\linewidth}
\begin{lstlisting}
def Teacher.L (teacher: Teacher Symbol): Language Symbol := {w: List Symbol | teacher.membershipQuery w}

def Teacher.Complete (teacher: Teacher Symbol) (n: Nat): Prop :=
  ∀ {State: Type} (ldfa: LearnedDFA Symbol State),
    ldfa.states.length ≤ n → teacher.equivalenceQuery ldfa = none → ldfa.accepts = teacher.L
\end{lstlisting}
\end{minipage}

Given a \lean{Teacher} instance that satisfies these guarantees, we can prove that the L*-algorithm terminates, and that its result is never larger than the minimal DFA of \lean{L}. 
For a complete teacher, it accepts exactly \lean{L}. 
To aid the termination proof, we define a function that gives the number of states in the minimal DFA for the teacher's language in the Listing below.
The number of states is equal to the number of equivalence classes of the Nerode congruence, which correspond one-to-one to the left quotients in mathlib.
Regularity guarantees that the set of left quotients is finite, so its cardinality is defined.
\lean{Language.IsRegular.finite_range_leftQuotient} states that the set of all possible left quotients for a regular language is finite.
With \lean{Set.Finite.toFinset} we convert this finite set into a \lean{Finset}, allowing us to access its cardinality with \lean{Finset.card}.
In mathlib, a \lean{Finset} is defined as a multiset without duplicates.

\begin{lstlisting}
noncomputable def Teacher.maxStates (teacher: Teacher Symbol): Nat :=
  teacher.prop_L.finite_range_leftQuotient.toFinset.card
\end{lstlisting}

\subsection{Implementing a Concrete Teacher}\label{sec:teacher-impl}
The \lean{Teacher} structure only specifies what a teacher has to provide.
To use it with the L*-algorithm, it must be instantiated with concrete queries and proofs of the required properties.
As an example, we consider a teacher for the language over the alphabet \lean{\{a,b\}} that accepts words with an even number of \lean{a}s and an even number of \lean{b}s, the running example of Angluin's paper~\cite{angluin_learning_1987}.
The minimal DFA for this language, \lean{ldfa}, is shown in Figure~\ref{fig:learned-dfa}.

The teacher answers membership queries by running \lean{ldfa}.
For equivalence queries, it tests random words: we use the random word teacher \lean{createRandomWordTeacher} described in Section~\ref{sec:random-word-teacher}, which takes the membership function, a proof that its language is regular, and the parameters of the random tests (a seed, the number of words, and their length).

\begin{lstlisting}
def membershipQuery (w: List Syms): Bool := decide (ldfa.eval w ∈ ldfa.accept)

theorem membershipQuery_eq_accepts: {w | membershipQuery w} = ldfa.accepts := by ...

def exampleTeacher: Teacher Syms := 
  createRandomWordTeacher membershipQuery (membershipQuery_eq_accepts ▸ ldfa.isRegular_accepts) 0 500 10
\end{lstlisting}

\begin{sloppypar}
The membership query decides whether the state reached by \lean{ldfa} is accepting\footnote{\leanfootnote{decide} turns this decidable proposition into a \leanfootnote{Bool}.}.
A teacher needs a proof that its language is regular.
The language of every \lean{LearnedDFA} is regular (\lean{LearnedDFA.isRegular_accepts}), since it is accepted by the DFA over the finite subtype of its states (Section~\ref{sec:learned-dfa-mathlib}).
The theorem \lean{membershipQuery_eq_accepts} transfers this to \lean{membershipQuery}\footnote{The operator \leanfootnote{$ $▸} rewrites the proof along this equation.}.
That every returned counterexample is genuine (\lean{prop_cex}) is proven once for \lean{createRandomWordTeacher}, so no further proof is needed here.
\end{sloppypar}

The resulting \lean{exampleTeacher} is sound but not complete.
Random tests may miss a difference between \lean{ldfa} and a hypothesis.
Hence, the termination and minimality results of Section~\ref{sec:lstar-proofs} apply to it, whereas correctness requires the additional assumption \lean{Teacher.Complete}.
When the target DFA is known, as here, a complete teacher is possible in principle: a search of the product automaton of \lean{ldfa} and the hypothesis finds a distinguishing word whenever one exists.
We leave such a teacher to future work.

\section{The L*-Algorithm}\label{sec:lstar-algo}

The core part of this work is the algorithm implementation itself, which takes a teacher as its input and produces a minimal \lean{LearnedDFA}.
For a complete teacher it represents exactly the unknown language.
The learning algorithm uses an observation table to keep track of observations from membership queries to the teacher.

The algorithm's learning is split into learning rounds. 
In Angluin's presentation, each round consists of (1) making the learner's observation table \textbf{closed} and \textbf{consistent}, (2) proposing a hypothesis DFA to the teacher through an equivalence query, and (3) processing a counterexample if one is returned (our implementation rotates this loop, see Section~\ref{sec:lstar-main}). 
After a learning round, the learner is either finished, or it has found at least one more state, making progress towards the minimal DFA known by the teacher.

This section details our formalization of this process. 
We will first describe the core data structures, the \lean{ObservationTable} and the \lean{Learner} that manages it, followed by the main algorithm loop. 
The formal proofs of the algorithm's termination, minimality, and correctness for complete teachers are presented in Section~\ref{sec:lstar-proofs}.

\subsection{The Observation Table}\label{sec:obs-table}
The central part of the learner in the L*-algorithm is an observation table, shown in the Listing below. 
We designed it as a structure with fields for the \lean{S} and \lean{E} sets, represented as lists to aid computability,
and a mathlib association list \lean{AList}\footnote{An association list, or \leanfootnote{AList} in mathlib, is a simple key-value map implemented as a list of \leanfootnote{(key, value)} pairs. It provides a lookup function to find the value associated with a given key, making it a straightforward, computable dictionary. \leanfootnote{\_} in the \leanfootnote{cache}'s type description in Lean is a placeholder for an unnamed parameter.} that caches membership-query results used to evaluate the function \lean{T}.

\begin{lstlisting}
structure ObservationTable (Symbol: Type) [alphabet: Alphabet Symbol] where
  S: List (List Symbol)
  E: List (List Symbol)
  cache: AList (fun (_: List Symbol) => Bool)

  prop_S: S.Nodup ∧ [] ∈ S ∧ ∀ s ∈ S, (∀ pre, pre <+: s → pre ∈ S)
  prop_E: E.Nodup ∧ [] ∈ E ∧ ∀ e ∈ E, (∀ suf, suf <:+ e → suf ∈ E)
  prop_cache: ... -- Guarantees cache covers all combinations from S, E, and A
\end{lstlisting}

The structure contains properties for its lists. 
\lean{prop_S} and \lean{prop_E} ensure that \lean{S} and \lean{E} have no duplicates, contain the empty word \lean{[]}, and are prefix-closed and suffix-closed, respectively. 
\lean{<+:} is the shorthand notation for "is prefix of", while \lean{<:+} means "is suffix of".
The \lean{prop_cache} property specifies that the cache contains entries for all words formed by concatenating elements from \lean{S} (and its one-symbol extensions) with elements from \lean{E}.

As seen in the Listing below, the function \lean{SA} produces all one-symbol extensions of all values in the \lean{S} set\footnote{In Lean the \leanfootnote{++} operator denotes List concatenation.}.
The function \lean{T} is implemented as a function backed by the cache.
The definition of an observation table \lean{row} is then based on \lean{T}, by iterating over all elements in the \lean{E} set.

\noindent\begin{minipage}{\linewidth}
\begin{lstlisting}
def ObservationTable.SA (obs: ObservationTable Symbol): List (List Symbol) :=
  obs.S.flatMap (fun s =>
    alphabet.symbols.map (fun a => s ++ [a])
  )

def ObservationTable.T (obs: ObservationTable Symbol): List Symbol → Bool := fun w =>
  match obs.cache.lookup w with
  | none => false  -- only reached if w is invalid (not queried yet)
  | some value => value

def ObservationTable.row (obs: ObservationTable Symbol) (s: List Symbol): List Bool :=
  obs.E.map (fun e => obs.T (s++e))
\end{lstlisting}
\end{minipage}

Each row in the observation table is a \lean{List Bool}. 
As the \lean{E} set is represented by a list, its elements have a natural order. 
Thus, if the \lean{row} function produces the same list for two input sequences, it means both sequences produce the same row values and are indistinguishable by the experiments in \lean{E}.
This is a key reason for choosing \lean{List} over \lean{Set}, as it simplifies row comparisons.

To better explain these concepts, we use a running example of learning the language of words with an even number of \lean{a}s and an even number of \lean{b}s. 
The initial observation table for this language is shown in Table~\ref{tab:obs-initial}. 
The \lean{S} and \lean{E} sets both start as \lean{[[]]} (a list containing only the empty sequence, $\epsilon$). 
The values in the table are the results of membership queries, e.g., \lean{membershipQuery([])} is \lean{true} and \lean{membershipQuery([a])} is \lean{false}. Hence,  $T(\epsilon) = 1$ and $T(a) = 0$ in Table~\ref{tab:obs-initial}.

\begin{table}[h]
    \centering
    \begin{tabular}{cc|c}
        \multicolumn{2}{c|} {\multirow{2}{*}{T}}                       & E          \\\cline{3-3}
        \multicolumn{2}{c|} {}                                         & $\epsilon$ \\\hline
        \multicolumn{1}{c|} {S}                           & $\epsilon$ & 1          \\\hline
        \multicolumn{1}{c|} {\multirow{2}{*}{S$\cdot$A}}  & a          & 0          \\
        \multicolumn{1}{c|} {}                            & b          & 0
    \end{tabular}
    \caption{Initial observation table for the even-even language.}
    \label{tab:obs-initial}
\end{table}

\subsubsection{Closed and Consistent Properties}\label{sec:obs-table-closed-consistent}

Before a \lean{LearnedDFA} can be constructed, the observation table must satisfy two properties:
\begin{itemize}
    \item An observation table is \textbf{closed} if every unique row in the extended \lean{S} set (\lean{S·A}) is already present in the \lean{S} set.
    \item An observation table is \textbf{consistent} if all input sequences in \lean{S} that produce equal rows also produce equal rows when extended with any symbol from the alphabet.
\end{itemize}

These properties are defined in Lean as follows:

\begin{lstlisting}
def ObservationTable.Closed (obs: ObservationTable Symbol) :=
  ∀ t ∈ obs.SA, ∃ s ∈ obs.S, obs.row s = obs.row t

def ObservationTable.Consistent (obs: ObservationTable Symbol) :=
  ∀ s₁ ∈ obs.S, ∀ s₂ ∈ obs.S, obs.row s₁ = obs.row s₂ →
  ∀ a, obs.row (s₁ ++ [a]) = obs.row (s₂ ++ [a])
\end{lstlisting}

To enforce these properties, we implement helper functions that search for sequences that would violate those properties.
If a violation is found, they return the prefix or suffix needed to resolve it.

Looking at Table~\ref{tab:obs-initial}, we can see that it is not closed. 
$row(a)$, which is $0$, does not match any row in the $S$ set (where the only row, for $\epsilon$, is $1$). 
We would fix this by moving a prefix with this new row (e.g., $a$) into the $S$ set, resulting in a table like Table~\ref{tab:obs-initial-closed}. 
This new table is now closed because the rows in $S\cdot A$ ($0$ for $b$ and $ab$, $1$ for $aa$) are all covered by rows in $S$ ($1$ for $\epsilon$, $0$ for $a$).

\begin{table}[h]
    \centering
    \begin{tabular}{cc|c}
        \multicolumn{2}{c|} {\multirow{2}{*}{T}}                       & E          \\\cline{3-3}
        \multicolumn{2}{c|} {}                                         & $\epsilon$ \\\hline
        \multicolumn{1}{c|} {\multirow{2}{*}{S}}          & $\epsilon$ & 1          \\
        \multicolumn{1}{c|} {}                            & a          & 0          \\\hline
        \multicolumn{1}{c|} {\multirow{3}{*}{S$\cdot$A}}  & b          & 0          \\
        \multicolumn{1}{c|} {}                            & aa         & 1          \\
        \multicolumn{1}{c|} {}                            & ab         & 0
    \end{tabular}
    \caption{Initial closed observation table for the even-even language.}
    \label{tab:obs-initial-closed}
\end{table}

Tables~\ref{tab:obs-initial} and~\ref{tab:obs-initial-closed} are both consistent, because the rows in \lean{S} are all unique.
In general, an L*-run may also reach a table like Table~\ref{tab:obs-table-inconsistent} (this particular table does not arise in our implementation, see Section~\ref{sec:learn-dfa}). 
This table is inconsistent. 
$row(a)$ and $row(b)$ are identical ($0$), but their continuations with $a$ differ: $row(aa)$ is $1$ while $row(ba)$ is $0$. 
We fix this by adding a as a distinguishing suffix to the $E$ set. 
This results in the table shown in Table~\ref{tab:obs-table-consistent}, which is now consistent because the rows in $S$ are all unique again.

\begin{figure}
\begin{minipage}[c]{0.48\linewidth}
    \centering
    \begin{tabular}{cc|c}
        \multicolumn{2}{c|} {\multirow{2}{*}{T}}                       & E          \\\cline{3-3}
        \multicolumn{2}{c|} {}                                         & $\epsilon$ \\\hline
        \multicolumn{1}{c|} {\multirow{3}{*}{S}}          & $\epsilon$ & 1          \\
        \multicolumn{1}{c|} {}                            & a          & 0          \\
        \multicolumn{1}{c|} {}                            & b          & 0          \\\hline
        \multicolumn{1}{c|} {\multirow{4}{*}{S$\cdot$A}}  & aa         & 1          \\
        \multicolumn{1}{c|} {}                            & ab         & 0          \\
        \multicolumn{1}{c|} {}                            & ba         & 0          \\
        \multicolumn{1}{c|} {}                            & bb         & 1
    \end{tabular}
    \captionof{table}{An inconsistent Observation Table.}
    \label{tab:obs-table-inconsistent}
\end{minipage}\hfill
\begin{minipage}[c]{0.48\linewidth}
    \centering
    \begin{tabular}{cc|cc}
        \multicolumn{2}{c|} {\multirow{2}{*}{T}}                       & \multicolumn{2}{c} {E} \\\cline{3-4}
        \multicolumn{2}{c|} {}                                         & $\epsilon$ & a         \\\hline
        \multicolumn{1}{c|} {\multirow{3}{*}{S}}          & $\epsilon$ & 1          & 0         \\
        \multicolumn{1}{c|} {}                            & a          & 0          & 1         \\
        \multicolumn{1}{c|} {}                            & b          & 0          & 0         \\\hline
        \multicolumn{1}{c|} {\multirow{4}{*}{S$\cdot$A}}  & aa         & 1          & 0         \\
        \multicolumn{1}{c|} {}                            & ab         & 0          & 0         \\
        \multicolumn{1}{c|} {}                            & ba         & 0          & 0         \\
        \multicolumn{1}{c|} {}                            & bb         & 1          & 0
    \end{tabular}
    \captionof{table}{A consistent Observation Table.}
    \label{tab:obs-table-consistent}
\end{minipage}
\end{figure}

\subsubsection{Creation of a DFA from an Observation Table}\label{sec:obs-table-create-dfa}

Once we have a closed and consistent observation table, we can produce a \lean{LearnedDFA}. 
A closed table ensures a complete DFA, while a consistent table ensures that the transitions are well-defined.

As shown in the Listing below\footnote{This listing uses Lean's pipeline operator \leanfootnotealt{|>}. \leanfootnotealt{x |> f} means applying the function \leanfootnote{f} to the value \leanfootnote{x}. For example, \leanfootnotealt{obs.S.map obs.row |> .dedup} is processed as \leanfootnote{List.dedup (List.map obs.row obs.S)}. \leanfootnote{List.dedup} removes duplicates from a list by performing a pairwise comparison.}, the components of the \lean{LearnedDFA} are derived directly from the table:
\begin{itemize}
    \item \lean{states}: List of unique rows from the S set.
    \item \lean{accept}: Rows of those words $s \in S$ with $T(s) = 1$.
    \item \lean{start}: State corresponding to the row of the empty sequence, \lean{row([])}.
    \item \lean{step}: Transition function that maps a state \lean{row(s)} and a symbol \lean{a} to the state \lean{row(s++[a])}.
\end{itemize}

\noindent\begin{minipage}{\linewidth}
\begin{lstlisting}
def ObservationTable.createDFA {obs: ObservationTable Symbol} (h_closed_consistent: obs.Closed ∧ obs.Consistent): LearnedDFA Symbol (List Bool) := 
  {
    states := obs.S.map obs.row |>.dedup
    accept := obs.S.filter obs.T |>.map obs.row |>.dedup
    start  := obs.row []
    step   := fun q a => ... -- Implements step based on q = row(s) and q' = row(s·a)
    ...  -- Proofs of well-formedness for the LearnedDFA
  }
\end{lstlisting}
\end{minipage}

Table~\ref{tab:obs-table-final} shows a possible final observation table for our example. 
From this table, we derive the \lean{LearnedDFA} depicted in Figure~\ref{fig:learned-dfa-final}. 
There are 4 states, corresponding to the unique rows $[1,0,0]$, $[0,1,0]$, $[0,0,1]$, and $[0,0,0]$. 
The state $[1,0,0]$ is the sole accepting state because it is the row of $\epsilon$ and $T(\epsilon) = 1$.
Transitions are determined by looking up the row of the extended prefix, i.e., the transition from state \lean{row([a])} on symbol $b$ is \lean{row([a,b])}.

\begin{figure}
\begin{minipage}[c]{0.48\linewidth}
    \centering
    \begin{tabular}{cc|cll}
        \multicolumn{2}{c|}{\multirow{2}{*}{T}}                       & \multicolumn{3}{c}{E} \\ \cline{3-5} 
        \multicolumn{2}{c|}{}                                         & $\epsilon$  & a  & b  \\ \hline
        \multicolumn{1}{c|}{\multirow{4}{*}{S}}          & $\epsilon$ & 1           & 0  & 0  \\
        \multicolumn{1}{c|}{}                            & a          & 0           & 1  & 0  \\
        \multicolumn{1}{c|}{}                            & b          & 0           & 0  & 1  \\
        \multicolumn{1}{c|}{}                            & ab         & 0           & 0  & 0  \\ \hline
        \multicolumn{1}{c|}{\multirow{5}{*}{S$\cdot$A}}  & aa         & 1           & 0  & 0  \\
        \multicolumn{1}{c|}{}                            & ba         & 0           & 0  & 0  \\
        \multicolumn{1}{c|}{}                            & bb         & 1           & 0  & 0  \\
        \multicolumn{1}{c|}{}                            & aba        & 0           & 0  & 1  \\
        \multicolumn{1}{c|}{}                            & abb        & 0           & 1  & 0
    \end{tabular}
    \captionof{table}{Possible final observation table for the even-even language.}
    \label{tab:obs-table-final}
\end{minipage}\hfill
\begin{minipage}[c]{0.48\linewidth}
    \centering
    \begin{tikzpicture}[node distance=2cm]
        \ltsnode{}{q0}{q0}
        \ltsnode{right of = q0}{q1}{q1}
        \ltsnode{below of = q0}{q2}{q2}
        \ltsnode{right of = q2}{q3}{q3}
        
        \node [circle, minimum height=.25cm,text width = .25cm, anchor=center,draw] at (q0.center){};
        
        \draw [arrow] ([yshift=2pt]q0.east) -- ([yshift=2pt]q1.west) node [midway, above] {\footnotesize$\mathsf{a}$};
        \draw [arrow] ([xshift=-2pt]q0.south) -- ([xshift=-2pt]q2.north) node [midway, left] {\footnotesize$\mathsf{b}$};
        \draw [arrow] ([yshift=-2pt]q1.west) -- ([yshift=-2pt]q0.east) node [midway, below] {\footnotesize$\mathsf{a}$};
        \draw [arrow] ([xshift=-2pt]q1.south) -- ([xshift=-2pt]q3.north) node [midway, left] {\footnotesize$\mathsf{b}$};
        \draw [arrow] ([yshift=2pt]q2.east) -- ([yshift=2pt]q3.west) node [midway, above] {\footnotesize$\mathsf{a}$};
        \draw [arrow] ([xshift=2pt]q2.north) -- ([xshift=2pt]q0.south) node [midway, right] {\footnotesize$\mathsf{b}$};
        \draw [arrow] ([yshift=-2pt]q3.west) -- ([yshift=-2pt]q2.east) node [midway, below] {\footnotesize$\mathsf{a}$};
        \draw [arrow] ([xshift=2pt]q3.north) -- ([xshift=2pt]q1.south) node [midway, right] {\footnotesize$\mathsf{b}$};
        
        \draw [arrow] node[auto] {} ++ (-1,0) to (q0);
    \end{tikzpicture}
    \caption{Visual representation of the \lean{LearnedDFA} generated from Table~\ref{tab:obs-table-final}.}
    \label{fig:learned-dfa-final}
\end{minipage}
\end{figure}

\subsection{The Learner}\label{sec:learner}
In our implementation, the \lean{Learner} is a structure that combines an observation table and a teacher. 
Its primary role is to manage the state of the observation table by querying the teacher and producing updated versions of itself.

\begin{lstlisting}
structure Learner (Symbol: Type) [alphabet: Alphabet Symbol] where
  obs: ObservationTable Symbol
  teacher: Teacher Symbol

  prop_obs_cache: ∀ w ∈ obs.cache, obs.cache.lookup w = some true ↔ w ∈ teacher.L
\end{lstlisting}

As shown in the Listing above, a \lean{Learner} contains an observation table and the teacher it learns from. 
The property \lean{prop_obs_cache} guarantees that all observations stored in the table's cache are consistent with the teacher's true language \lean{L}.

Due to the nature of purely functional programs where no side effects are allowed, the \lean{Learner} structure is constantly replaced by an updated version containing changes to its observation table. 
This is achieved through the following key operations.

\subsubsection{Learner Initialization}\label{sec:learner-init}
When starting the L*-algorithm, we first create an initial \lean{Learner}. 
The \lean{init} function (see Listing below) constructs an observation table with the \lean{S} and \lean{E} sets both containing only the empty sequence. 
It then populates the initial \lean{cache} by making membership queries to the teacher for the empty sequence and all single-symbol sequences.
For our running example, this \lean{init} function is exactly what produces the initial table shown previously in Table~\ref{tab:obs-initial}.

\noindent\begin{minipage}{\linewidth}
\begin{lstlisting}
def Learner.init (teacher: Teacher Symbol): Learner Symbol :=
  let cache := ...  -- Create cache containing all single symbol words
  {
    obs := ObservationTable.init cache (by ...)
    teacher := teacher
    prop_obs_cache := by ...
  }
\end{lstlisting}
\end{minipage}

\subsubsection{Counterexample Processing}\label{sec:learner-cex-proc}
One major part of the L*-algorithm is counterexample processing. 
If an equivalence query is answered with a counterexample \lean{cex} by the teacher, the learner takes \lean{cex} and its prefix-closed set and adds them to the \lean{S} set.

As shown in the Listing below, the function first calculates the new prefixes to be added. 
It then extends the \lean{cache} with all the new membership queries required by the expanded \lean{S} set and constructs a new \lean{Learner} with the updated observation table. 
For example, if the learner built a faulty \lean{LearnedDFA} and the teacher returned \lean{cex = [a,b]}, \lean{processCex} would add \lean{[a]} and \lean{[a,b]} to the S set (if not already present) and query the teacher for new words like \lean{[a,a]}, \lean{[a,b]}, \lean{[a,b,a]}, and \lean{[a,b,b]} to fill in the new table entries.

\begin{lstlisting}
def Learner.processCex (learner: Learner Symbol) (cex: List Symbol): Learner Symbol :=
  let addedRows := prefixClosedSet cex |>.filter (· ∉ learner.obs.S)
  let cache := ... -- Extend cache with new queries for addedRows
  let obs: ObservationTable Symbol := {learner.obs with
    S := learner.obs.S ++ addedRows
    cache := cache
    prop_S := by ...
    prop_cache := by ...
  }
  {learner with
    obs := obs
    prop_obs_cache := by ...
  }
\end{lstlisting}

\subsubsection{Producing a Closed and Consistent Observation Table}\label{sec:learner-closed-consistent}
The second major part of a learning round is making the observation table closed and consistent. 
This is a stepwise, recursive process that repeats until the table converges. 
The \lean{makeClosedConsistent} function in the Listing below orchestrates this by first fixing any inconsistencies, and then fixing any unclosed rows recursively.

\begin{lstlisting}
def Learner.makeClosedConsistent (learner: Learner Symbol): Learner Symbol :=
  match h_inconsistent_suffix: learner.obs.findInconsistentSuffix with
  | some _ => (makeConsistentStep h_inconsistent_suffix).makeClosedConsistent
  | none => match h_unclosed_row: learner.obs.findUnclosedRow with
    | some _ => (makeClosedStep h_unclosed_row).makeClosedConsistent
    | none => learner
\end{lstlisting}

Using our running example, we can trace the function's logic as follows:
\begin{itemize}
    \item Fixing an unclosed table:
    When called with a learner whose table is the initial one (Table~\ref{tab:obs-initial}), \lean{findInconsistentSuffix} returns none.
    However, \lean{findUnclosedRow} finds a violation, as the \lean{row([a])} ($0$) is not in the \lean{S} set. 
    It returns \lean{some [a]}. 
    The algorithm then calls a helper function, \lean{makeClosedStep}, which adds \lean{[a]} to the \lean{S} set, queries the teacher for new words like \lean{[a,a]} and \lean{[a,b]}, and recursively calls \lean{makeClosedConsistent}. 
    This action is what transforms Table~\ref{tab:obs-initial} into Table~\ref{tab:obs-initial-closed}.

    \begin{sloppypar}
    \item Fixing an inconsistent table:
    An L*-run may reach an observation table like Table~\ref{tab:obs-table-inconsistent}. 
    This time, \lean{findInconsistentSuffix} finds a violation: \lean{row([a])} and \lean{row([b])} are equal, but their continuations with \lean{a} are not. 
    It returns \lean{some [a]}.
    The algorithm then calls \lean{makeConsistentStep}, which adds \lean{[a]} to the \lean{E} set, queries the teacher for all the new experiments, i.e., \lean{[a,a,a]}, \lean{[a,b,a]}, \lean{[b,a,a]}, and \lean{[b,b,a]}, and recursively calls \lean{makeClosedConsistent}.
    This action transforms Table~\ref{tab:obs-table-inconsistent} into the consistent Table~\ref{tab:obs-table-consistent}.
    \end{sloppypar}
\end{itemize}

This process repeats, first resolving all inconsistencies, then resolving an unclosed row, until both \lean{findInconsistentSuffix} and \lean{findUnclosedRow} return \lean{none}, at which point the final, valid \lean{Learner} is returned.
The final \lean{Learner} then holds a closed and consistent observation table.

\subsection{The Main L*-Algorithm}\label{sec:lstar-main}
The \lean{lstar} algorithm, shown in the Listing below, is the top-level function that takes a \lean{Teacher} instance and returns the final hypothesis. 
It is minimal, and correct for a complete teacher.

First, in Line~\ref{ln:learner-init} we initialize a \lean{Learner}, which creates an initial observation table. 
In Line~\ref{ln:call-lstar-aux} we call \lean{lstarAux}, the recursive implementation of the learning rounds.
This takes a \lean{Learner}, with a closed and consistent observation table.

\begin{sloppypar}
\lean{learner.makeClosedConsistent_closed_consistent} is a proof that the table of \lean{learner.makeClosedConsistent} is closed and consistent.
As explained later, Lean automatically infers that the \lean{Learner} to use for \lean{lstarAux} is \lean{learner.makeClosedConsistent}.
Consequently, this ensures that the table is closed and consistent before a learning round begins.
\end{sloppypar}

\begin{lstlisting}
def lstar (teacher: Teacher Symbol): LearnedDFA Symbol (List Bool) :=
  let learner := Learner.init teacher(*@\label{ln:learner-init}@*)
  lstarAux learner.makeClosedConsistent_closed_consistent(*@\label{ln:call-lstar-aux}@*)
\end{lstlisting}

The recursive \lean{lstarAux} function (see simplified Listing below) performs a single learning round. 
It takes a \lean{Learner} whose observation table is guaranteed by the type system to be closed and consistent.

\begin{lstlisting}
def lstarAux {learner: Learner Symbol} (h_closed_consistent: learner.obs.Closed ∧ learner.obs.Consistent): LearnedDFA Symbol (List Bool) :=
  let ldfa := learner.obs.createDFA h_closed_consistent
  match learner.teacher.equivalenceQuery ldfa with
  | none => ldfa
  | some cex => lstarAux (learner.processCex cex).makeClosedConsistent_closed_consistent
\end{lstlisting}

Each learning round in our implementation starts with a learner holding a closed and consistent observation table.
Then, a learning round progresses as follows:
\begin{enumerate}
    \item We create a hypothesis \lean{LearnedDFA}, \lean{ldfa}, from the learner's observation table (using the \lean{createDFA} function discussed in Section~\ref{sec:obs-table-create-dfa}).
    \item We then present this \lean{ldfa} to the teacher via an equivalence query.
    \item If the teacher responds with
    \begin{itemize}
        \item \lean{none}, the teacher found no counterexample. 
        We return \lean{ldfa} (which is correct if the teacher is complete).
        \item \lean{some cex}, we process this counterexample with \lean{processCex}, which extends the \lean{S} set.
    \end{itemize}
    \item Finally, if we got a counterexample, we call \lean{makeClosedConsistent} on the new learner to make the observation table closed and consistent again before starting the next learning round by recursively calling \lean{lstarAux}.
\end{enumerate}

The curly braces for the parameter \lean{\{learner: ...\}} denote an implicit argument. 
This allows Lean to automatically infer which \lean{Learner} instance is being used from the type of the second parameter, \lean{h_closed_consistent}, which is a proof that the learner's observation table is closed and consistent. 
This mechanism allows us to pass both the data (\lean{learner}) and a proof about it into the next recursive call in a clean and type-safe manner.

This recursive loop is the core that runs the learning process. 
It orchestrates the \lean{Learner} operations to transform the observation table through the stages shown in our running example, from the initial state (Table~\ref{tab:obs-initial}), through various intermediate states (Table~\ref{tab:obs-initial-closed}), until a final hypothesis can be built from a table (Table~\ref{tab:obs-table-final}).
The proofs that this process always terminates, yields a minimal DFA, and yields the correct one for a complete teacher will be detailed in Section~\ref{sec:lstar-proofs}.

\section{Machine-Checked Proofs of the L*-Algorithm}\label{sec:lstar-proofs}

Using the definitions and implementation from Sections~\ref{sec:formalizing-dfas},~\ref{sec:teacher}, and~\ref{sec:lstar-algo}, we now present the machine-checked proofs about our L*-algorithm.
The proofs follow the structure of Angluin's original arguments but are constructed within Lean, guaranteeing their soundness.

The section is split into three parts:
\begin{itemize}
    \item Partial correctness on observation tables (Section~\ref{sec:partial-correctness}): A DFA constructed from a closed and consistent observation table is consistent with the table and minimal among all DFAs that are consistent with it (Theorem~\ref{thm:minimality}).
    \item Termination (Section~\ref{sec:termination}): For every teacher whose language is regular and whose counterexamples are genuine, the algorithm terminates.
    \item Correctness, minimality, and uniqueness of the result (Section~\ref{sec:lstar-results}): The learned DFA is never larger than the minimal DFA of the teacher's language. 
    For a complete teacher, the learned DFA accepts exactly that language and is the minimal DFA, unique up to isomorphism.
\end{itemize}

\subsection{Partial Correctness of the L*-Algorithm}\label{sec:partial-correctness}
In Angluin's argument, partial correctness means that whenever L* stops, its hypothesis agrees with all observations and is minimal among the acceptors that do.
That it accepts the target language then follows from the teacher's answer.
In our formalization, this requires a complete teacher (Section~\ref{sec:lstar-results}).
The arguments and lemmas presented in this section are a direct formalization of those originally presented by Angluin~\cite{angluin_learning_1987}.

We show that the DFA produced from any closed, consistent observation table is the minimal DFA, unique up to isomorphism, that is consistent with the information in that table.
This property is captured by the following theorem.

\begin{theorem}[\cite{angluin_learning_1987}]\label{thm:minimality}
If (S, E, T) is a closed, consistent observation table, then the acceptor M(S, E, T) is consistent with the finite function T. 
Any other acceptor consistent with T but inequivalent to M(S, E, T) must have more states.
\end{theorem}

Our main goal for this subsection is to prove the Lean formalization of this theorem, shown in the Listing below. 
All results of this section are stated for an observation table.
They do not depend on the teacher.
For the proof, we follow the structure of the original L*-algorithm paper, splitting it into a series of formal lemmas.
\begin{lstlisting}
theorem ObservationTable.createDFA_minimal {obs: ObservationTable Symbol} (h_closed_consistent: obs.Closed ∧ obs.Consistent):
  let dfa: DFA Symbol (List Bool) := obs.createDFA h_closed_consistent;
  obs.ConsistentWith dfa ∧
  (∀ (DFAState: Type) [DecidableEq DFAState] [Fintype DFAState] (otherDFA: DFA Symbol DFAState),
    obs.ConsistentWith otherDFA ∧ dfa.accepts ≠ otherDFA.accepts →
    Fintype.card DFAState > obs.nStates)
\end{lstlisting}

Central to this theorem is the property \lean{ObservationTable.ConsistentWith}. 
It formalizes that for a DFA to be consistent with an observation table, its language must agree with the table's \lean{T} function for all known prefixes and experiments. 
We define it as follows:
\begin{lstlisting}
def ObservationTable.ConsistentWith (obs: ObservationTable Symbol) (dfa: DFA Symbol State) :=
  ∀ s ∈ (obs.S ++ obs.SA), ∀ e ∈ obs.E, obs.T (s ++ e) ↔ (s ++ e) ∈ dfa.accepts
\end{lstlisting}
Additionally, \lean{DFA.accepts} describes the language of a mathlib DFA, \lean{Fintype.card DFAState} the number of states of \lean{otherDFA}, and \lean{obs.nStates}\footnote{\leanfootnote{ObservationTable.nStates} returns the number of unique rows in an observation table.} the number of states of the learned DFA.
We represent \textit{any other acceptor} from the theorem by a mathlib DFA whose state type has a \lean{Fintype} instance, such that it has finitely many states, and a \lean{DecidableEq} instance required by the proof of Lemma~\ref{thm:consistent-dfa-isomorphic} to compare states.

A \lean{Fintype} instance states that a type has finitely many elements.
It contains a finite set \lean{elems} and a proof that every element of the type is in it (see the Listing below). 
\lean{Fintype.card} is the number of these elements.
For the state type of a DFA, it is the number of states.
\begin{lstlisting}
class Fintype (α : Type*) where
  elems : Finset α  -- The `Finset` containing all elements of a `Fintype`
  complete : ∀ x : α, x ∈ elems  -- A proof that `elems` contains every element of the type
\end{lstlisting}

The first foundational lemma states that for any sequence $s$ in $S$ or $S\cdot A$, the state reached in the learned DFA is precisely $row(s)$.
\begin{lemma}[\cite{angluin_learning_1987}]\label{thm:eval-eq-row}
Assume that (S, E, T) is a closed, consistent observation table. 
For the acceptor M(S, E, T) and for every $s$ in ($S\cup S \cdot A$), $\delta(q_0,s) = row(s)$.
\end{lemma}

We prove the formalization shown in the Listing below by induction over the input sequence.
The inductive step works because the \lean{step} function of a \lean{LearnedDFA} is constructed from \mbox{\lean{row(s ++ [a])}}, allowing us to deconstruct the input sequence symbol by symbol.
\begin{lstlisting}
theorem ObservationTable.createDFA_eval_eq_row {obs: ObservationTable Symbol} (h_closed_consistent: obs.Closed ∧ obs.Consistent):
  let dfa := obs.createDFA h_closed_consistent;
  ∀ s, s ∈ obs.S ∨ s ∈ obs.SA → dfa.eval s = obs.row s
\end{lstlisting}

Next, we establish that the learned DFA is consistent with the observation table's \lean{T} function.
\begin{lemma}[\cite{angluin_learning_1987}]\label{thm:learned-dfa-consistent-with-T}
Assume that (S, E, T) is a closed, consistent observation table. 
Then the acceptor M(S, E, T) is consistent with the finite function T. 
That is, for every $s$ in ($S\cup S\cdot A$) and $e$ in $E$, $\delta(q_0,s\cdot e)$ is in $F$ if and only if~$T(s\cdot e) = 1$.
\end{lemma}

This lemma, formalized in the Listing below, is a direct consequence of how \lean{createDFA} works, as it uses the function \lean{T} to define the \lean{LearnedDFA}'s accepting states (see Section~\ref{sec:obs-table-create-dfa}).
\begin{lstlisting}
theorem ObservationTable.createDFA_consistentWith {obs: ObservationTable Symbol} (h_closed_consistent: obs.Closed ∧ obs.Consistent):
  let dfa: DFA Symbol (List Bool) := obs.createDFA h_closed_consistent;
  obs.ConsistentWith dfa
\end{lstlisting}

With these preliminaries, the core of the minimality proof rests on the next two lemmas. 
First, we prove that any DFA consistent with the observation table must have at least as many states as there are unique rows in \lean{S}.
\begin{lemma}[\cite{angluin_learning_1987}]\label{thm:consistent-dfa-min-states}
Let (S, E, T) be an observation table. 
Let n denote the number of different values of row(s) for s in S. 
Any acceptor consistent with T must have at least n states.
\end{lemma}

The proof constructs an injective map from the distinct rows of $S$ to the states of any consistent DFA, mapping $row(s)$ to the state the DFA reaches with $s$.
Our Lean formalization shown in the Listing below requires instances for \lean{DecidableEq} and \lean{Fintype} on the state type to ensure comparability and to reason about cardinality.
\begin{lstlisting}
theorem ObservationTable.nStates_le_card_of_consistentWith [DecidableEq DFAState] [Fintype DFAState] (obs: ObservationTable Symbol) (dfa: DFA Symbol DFAState):
  obs.ConsistentWith dfa → obs.nStates ≤ Fintype.card DFAState
\end{lstlisting}

Second, we prove that if another consistent DFA has the same number of states (or fewer), it must be isomorphic to the one we constructed.
\begin{lemma}[\cite{angluin_learning_1987}]\label{thm:consistent-dfa-isomorphic}
Assume that (S, E, T) is a closed, consistent observation table. 
Suppose the acceptor M(S, E, T) has n states. 
If M' = (Q',$q_0$',F',$\delta$') is any acceptor consistent with T that has n or fewer states, then M' is isomorphic to M(S, E, T).
\end{lemma}

Our formalization in the Listing below constructs this isomorphism. 
To do so, we introduce a generic state type \lean{DFAState} with \lean{DecidableEq} and \lean{Fintype} instances, for comparability and finiteness.
A key step is using \lean{LearnedDFA.toAttachedDFA} to get a mathlib \lean{DFA} representation of the \lean{LearnedDFA} whose state type consists only of its listed states (the rows of $S$), which is necessary for defining a bijection \lean{$ $Φ}\footnote{In mathlib \leanfootnote{$ $≃} is the notation for type equivalence, i.e., a bijection between two types. \leanfootnote{$ $Φ} represents the forward function of the bijection.} between the state sets.
We then prove that \lean{$ $Φ} preserves the initial state, accepting states, and the step function.

\begin{lstlisting}
theorem ObservationTable.iso_of_consistentWith_of_card_le (obs: ObservationTable Symbol) (h_closed_consistent: obs.Closed ∧ obs.Consistent):
  let dfa := (obs.createDFA h_closed_consistent).toAttachedDFA;
  let n := obs.nStates;
  ∀ (DFAState: Type) [DecidableEq DFAState] [Fintype DFAState] (otherDFA: DFA Symbol DFAState),
    obs.ConsistentWith otherDFA ∧ Fintype.card DFAState ≤ n →
  (∃ Φ: {q // q ∈ obs.states} ≃ DFAState,
    Φ dfa.start = otherDFA.start ∧
    (∀ q, q ∈ dfa.accept ↔ Φ q ∈ otherDFA.accept) ∧
    (∀ q, ∀ a, Φ (dfa.step q a) = otherDFA.step (Φ q) a))
\end{lstlisting}

These last two lemmas work in tandem. 
Lemma~\ref{thm:consistent-dfa-min-states} establishes that any consistent DFA must have at least n states (the number of unique rows).
Lemma~\ref{thm:consistent-dfa-isomorphic} then shows that if a consistent DFA has exactly n states, it must be isomorphic to our learned DFA. 
Together, they prove that the learned DFA is the minimal automaton for the information contained in the observation table, unique up to isomorphism, thus proving Theorem~\ref{thm:minimality}.

\subsection{Termination of the L*-Algorithm}\label{sec:termination}
Next, we prove that \lean{lstar} terminates.
As our algorithm is implemented recursively, we need to show that it converges to a fixed point, i.e., the algorithm halts when the teacher's equivalence query produces no counterexample.

In Lean, proving termination for a recursive function requires specifying a measure that strictly decreases with every recursive call.
The L*-algorithm's progress is marked by the increasing number of states in the learned hypothesis, \lean{nStates}, which grows towards a finite upper bound, the number of states in the minimal target DFA \lean{maxStates}. 
Lean cannot automatically infer a decreasing measure from this increasing property, so we must provide one manually. 
The termination measure is the difference between \lean{maxStates} and \lean{nStates}:

\begin{lstlisting}
termination_by learner.teacher.maxStates - learner.obs.nStates
\end{lstlisting}

To satisfy Lean's termination checker, we must prove that this measure strictly decreases.
Since \lean{maxStates} is a constant and \lean{nStates} never exceeds it (Section~\ref{sec:myhill-nerode-bound}), it suffices to prove that \lean{nStates} strictly increases with every recursive call.
A recursive call only occurs when an equivalence query fails and returns a counterexample. 
Therefore, our proof obligation is to show that processing a counterexample guarantees that at least one new state is added to the hypothesis DFA.

\subsubsection{Proving Progress in a Learning Round}
We can demonstrate this progress from two complementary perspectives.
The termination proof of \lean{lstarAux} uses the high-level argument.
The low-level one is formalized as a separate theorem.

\paragraph{High-Level Proof using Minimality}
This proof, used by Angluin~\cite{angluin_learning_1987} and in the termination proof of our Lean implementation, reuses the minimality theorem (Theorem~\ref{thm:minimality}) from the partial correctness argument.
The logic is as follows:
\begin{enumerate}
    \item Let \lean{dfa} be the \lean{LearnedDFA} generated from the closed, consistent observation table just before the equivalence query, with \lean{n} states.
    
    \item The equivalence query fails, yielding a counterexample \lean{cex}. 
    This means the language of \lean{dfa} is not the target language (this needs only \lean{prop_cex}).
    
    \item The counterexample \lean{cex} and its prefixes are added to the observation table. 
    The table is then made closed and consistent again, producing a new DFA, \lean{dfa'}, with \lean{n'} states.
    
    \item By construction, \lean{dfa'} correctly classifies \lean{cex}, while \lean{dfa} does not. 
    Therefore, the languages of \lean{dfa} and \lean{dfa'} are inequivalent (as \lean{cex} is in $S$ and $\epsilon$ in $E$, and all cached values agree with the teacher).
    
    \item Crucially, the process of adding a counterexample and re-establishing the table's properties only adds information.
    Existing observations are never invalidated. 
    Thus, \lean{dfa'} remains consistent with the original observation table used to generate \lean{dfa}.
    
    \item From Theorem~\ref{thm:minimality}, we know that any DFA (including \lean{dfa'}) that is consistent with the original table but inequivalent to \lean{dfa} must have more states.
    
    \item Therefore, \lean{n' > n}. 
    The number of states strictly increases.
\end{enumerate}

\paragraph{Low-Level Proof via Table Properties}
This proof examines the effect of a counterexample on the observation table itself.
Processing a genuine counterexample always makes the table inconsistent.
Assume the contrary.
Then, by induction over all prefixes $p$ of \lean{cex}, the row of $p$ in the new table equals the state $\delta(q_0, p)$ that the old hypothesis reaches with $p$.
For the empty prefix this holds because existing observations remain unchanged.
For $p\cdot a$, closedness of the old table provides a representative $s \in S$ with $row(s) = \delta(q_0, p)$, consistency of the new table transfers $row(p) = row(s)$ to $row(p\cdot a) = row(s\cdot a)$, and $row(s\cdot a) = \delta(q_0, p\cdot a)$ by Lemma~\ref{thm:eval-eq-row}.
In particular, the row of \lean{cex} in the new table is the state the old hypothesis reaches with \lean{cex}.
Comparing their entries for the experiment $\epsilon$ shows that the old hypothesis classifies \lean{cex} like the teacher, which contradicts \lean{cex} being a counterexample.

\noindent\begin{minipage}{\linewidth}
\begin{lstlisting}
theorem Learner.processCex_not_consistent (learner: Learner Symbol) (h_closed_consistent: learner.obs.Closed ∧ learner.obs.Consistent) (cex: List Symbol) (h_cex: ¬(cex ∈ (learner.obs.createDFA h_closed_consistent).accepts ↔ cex ∈ learner.teacher.L)):
  ¬(learner.processCex cex).obs.Consistent

theorem Learner.nStates_lt_makeClosedConsistent_processCex (learner: Learner Symbol) (h_closed_consistent: learner.obs.Closed ∧ learner.obs.Consistent) (cex: List Symbol) (h_cex: ¬(cex ∈ (learner.obs.createDFA h_closed_consistent).accepts ↔ cex ∈ learner.teacher.L)):
  learner.obs.nStates < (learner.processCex cex).makeClosedConsistent.obs.nStates
\end{lstlisting}
\end{minipage}

Since the table is inconsistent, the subsequent call to \lean{makeClosedConsistent} increases the number of states, as each of its steps increases the number of distinct rows of $S$:
\begin{itemize}
    \item If the table is not closed:
    There exists some \lean{s} in the new \lean{S} set and \lean{a} in the alphabet such that \lean{row(s ++ [a])} is not equal to \lean{row(s')} for any \lean{s'} in \lean{S}. 
    The \lean{makeClosedStep} moves \lean{s ++ [a]} into \lean{S}, adding a new, unique row and thus increasing \lean{nStates}.

    \begin{sloppypar}
    \item If the table is not consistent:
    There exist \lean{s₁}, \lean{s₂} in \lean{S} such that \lean{row(s₁) = row(s₂)}, but for some symbol \lean{a} and \lean{e} in \lean{E}, \lean{T(s₁ ++ [a] ++ e) ≠ T(s₂ ++ [a] ++ e)}.
    The \lean{makeConsistentStep} will add the suffix \lean{[a] ++ e} to \lean{E} which separates the rows of \lean{s₁} and \lean{s₂}.
    This breaks the equality \lean{row(s₁) = row(s₂)}, creating at least one new unique row in \lean{S} and thus increasing \lean{nStates}.
    \end{sloppypar}
\end{itemize}

In all cases, processing a counterexample and restoring the table's properties guarantees that \lean{nStates} strictly increases as shown in the Listing above.

\subsubsection{The Myhill-Nerode Bound}\label{sec:myhill-nerode-bound}
The final piece of the termination proof is to show that the number of states, \lean{nStates}, has a finite upper bound \lean{maxStates}. 
\lean{maxStates} is provided by the Myhill-Nerode theorem, which states that any regular language has a finite number of equivalence classes, i.e., left quotients, \lean{Language.leftQuotient}, in mathlib.
The states of the minimal DFA for a language correspond directly to these classes.

We prove that the number of distinct rows in an observation table (which corresponds to \lean{nStates}) can never exceed the number of Myhill-Nerode classes of the teacher's language. 
This is formalized through two key lemmas. 
First, we show that if two prefixes fall into the same Myhill-Nerode class, they must produce the same row in the observation table.
\begin{lstlisting}
theorem Learner.row_eq_of_leftQuotient_eq (learner: Learner Symbol):
  ∀ s₁ ∈ learner.obs.S, ∀ s₂ ∈ learner.obs.S,
  learner.teacher.L.leftQuotient s₁ = learner.teacher.L.leftQuotient s₂ → learner.obs.row s₁ = learner.obs.row s₂
\end{lstlisting}

This implies that the mapping from a Myhill-Nerode class to an observation table row is well-defined. 
From this, we can conclude that the number of unique rows is less than or equal to the number of unique Myhill-Nerode classes represented in the S set.\footnote{\leanfootnote{Set.image}, similar to \leanfootnote{List.map}, applies a function to every element in a mathlib set. \leanfootnote{Set.ncard} returns the cardinality of a mathlib set.}
\begin{lstlisting}
theorem Learner.ncard_rows_le_ncard_leftQuotients (learner: Learner Symbol):
  ({s | s ∈ learner.obs.S}.image (learner.obs.row)).ncard ≤
  ({s | s ∈ learner.obs.S}.image (learner.teacher.L.leftQuotient)).ncard
\end{lstlisting}

Since the number of Myhill-Nerode classes for the entire language is an upper bound on the number of classes for any subset of its prefixes, the number of states in the minimal DFA for the teacher's language serves as a finite upper bound for \lean{nStates}. 
This is the \lean{maxStates} value used in our termination measure:

\begin{lstlisting}
theorem Learner.nStates_le_maxStates (learner: Learner Symbol):
  learner.obs.nStates ≤ learner.teacher.maxStates
\end{lstlisting}

\subsubsection{Termination Argument}
We have now established both necessary conditions for termination:
\begin{itemize}
    \item Strict Progress: Each recursive call to \lean{lstarAux}, which only happens after processing a counterexample, strictly increases the number of states (\lean{nStates}) in the hypothesis DFA.

    \item Finite Bound: The number of states is bounded by the number of Myhill-Nerode classes of the target language, a finite number.
\end{itemize}

Since a strictly increasing natural number cannot grow indefinitely within a finite bound, the number of learning rounds must be finite.
Therefore, the \lean{lstar} algorithm is guaranteed to terminate.
Section~\ref{sec:lstar-results} uses this to state the guarantees for the result of \lean{lstar}.

\subsection{Correctness, Minimality, and Uniqueness of the Learned DFA}\label{sec:lstar-results}
Theorem~\ref{thm:minimality} and the termination proof are statements about observation tables and learning rounds.
To state guarantees for the result of \lean{lstar}, we first characterize what \lean{lstarAux} returns: the hypothesis of some closed and consistent learner with the same teacher, for which the teacher returns no counterexample.

\paragraph{Correctness}
The recursion stops when the teacher returns \lean{none}.
For an incomplete teacher, e.g., the random-word teacher detailed in Section~\ref{sec:random-word-teacher}, this does not imply that the learned DFA is correct.
Correctness therefore requires a complete teacher:

\begin{lstlisting}
theorem lstar_correct (teacher: Teacher Symbol) (h_complete: teacher.Complete teacher.maxStates):
  (lstar teacher).accepts = teacher.L
\end{lstlisting}

By the definition of \lean{lstarAux}, the teacher returns \lean{none} for the result, and by the Myhill-Nerode bound (Section~\ref{sec:myhill-nerode-bound}), the result has at most \lean{teacher.maxStates} states.
Consequently, completeness for hypotheses of this size implies correctness of the result.
This is why \lean{Teacher.Complete} is parameterized by a number of states.
L* never proposes a hypothesis with more than \lean{teacher.maxStates} states, so completeness is only needed up to this bound.

\paragraph{Minimality}
Minimality holds for every teacher, even for incomplete ones:

\begin{lstlisting}
theorem lstar_minimal (teacher: Teacher Symbol) {DFAState: Type} [Fintype DFAState] (dfa: DFA Symbol DFAState) (h_dfa_accepts: dfa.accepts = teacher.L):
  (lstar teacher).states.length ≤ Fintype.card DFAState
\end{lstlisting}

The result has at most \lean{teacher.maxStates} states, and every DFA accepting the teacher's language has at least \lean{teacher.maxStates} states, as each left quotient of the language is represented by one of its states.

\paragraph{Uniqueness}
Every DFA accepting the teacher's language is consistent with the final observation table, as all observations during learning agree with the teacher.
We can use Lemma~\ref{thm:consistent-dfa-isomorphic} to prove uniqueness of the result.

\noindent\begin{minipage}{\linewidth}
\begin{lstlisting}
theorem lstar_unique (teacher: Teacher Symbol) {DFAState: Type} [DecidableEq DFAState] [Fintype DFAState] (dfa: DFA Symbol DFAState) (h_dfa_accepts: dfa.accepts = teacher.L) (h_card_le: Fintype.card DFAState ≤ (lstar teacher).states.length):
  let learnedDFA := (lstar teacher).toAttachedDFA;
  ∃ Φ: {state // state ∈ (lstar teacher).states} ≃ DFAState,
    Φ learnedDFA.start = dfa.start ∧
    (∀ state, state ∈ learnedDFA.accept ↔ Φ state ∈ dfa.accept) ∧
    (∀ state, ∀ a, Φ (learnedDFA.step state a) = dfa.step (Φ state) a)

theorem lstar_states_length_eq_maxStates (teacher: Teacher Symbol) (h_complete: teacher.Complete teacher.maxStates):
  (lstar teacher).states.length = teacher.maxStates
\end{lstlisting}
\end{minipage}

\begin{sloppypar}
For a complete teacher, the result has exactly the number of states of the minimal DFA \lean{teacher.maxStates}.
Together with \lean{lstar_correct} and \lean{lstar_unique}, the learned DFA is the minimal DFA of the teacher's language, unique up to isomorphism.
For an incomplete teacher whose result is wrong, \lean{lstar_unique} holds only vacuously: the result has fewer than \lean{teacher.maxStates} states, so no DFA accepting the teacher's language satisfies \lean{h_card_le}.
\end{sloppypar}

\section{Running the L*-Algorithm Implementation}\label{sec:run-lstar}

Our implementation is not only a mathematical specification of the L*-algorithm, it also serves as a computable implementation. 
Additionally, attached to this implementation are proofs of termination, minimality, and uniqueness for every teacher, and correctness for every complete teacher, as described in Section~\ref{sec:lstar-proofs}.

One of our main goals was to provide a straightforward, easy-to-use interface, where we can leverage the proofs on top of the implementation if needed.
Therefore, we chose the following signature for the implementation:

\begin{lstlisting}
def lstar (teacher: Teacher Symbol): LearnedDFA Symbol (List Bool) := ...
\end{lstlisting}

We provide the algorithm with a \lean{Teacher} to learn the unknown regular language from, and it produces a \lean{LearnedDFA}.
This structure provides list representations of its states and accepting states for use in computable contexts, and translations to a mathlib \lean{DFA} for use in other proofs.

\subsection{Random Word Teacher Implementation}\label{sec:random-word-teacher}

Before running the Lean implementation of the L*-algorithm, we first need a teacher to learn from.
As described in Section~\ref{sec:teacher}, every \lean{Teacher} must guarantee two properties:

\begin{enumerate}
    \item The unknown language, defined by the membership query, is regular (\lean{prop_L}).
    \item Every counterexample returned by the equivalence query distinguishes the hypothesis from the unknown language, i.e., exactly one of the two accepts it (\lean{prop_cex}).
\end{enumerate}

These properties suffice for termination, minimality, and uniqueness.
Correctness additionally requires a complete teacher (\lean{Teacher.Complete}), whose equivalence query returns \lean{none} only for a correct hypothesis.

Here we present a random word teacher using Lean's built-in random number generator \lean{mkStdGen}, shown in the Listing below.
It provides both guarantees, but it is not complete.

\noindent\begin{minipage}{\linewidth}
\begin{lstlisting}
def createRandomWordTeacher (membershipQuery: List Symbol → Bool) (h_target_regular: Language.IsRegular {w: List Symbol | membershipQuery w}) (seed: Nat) (nWords: Nat) (wordLen: Nat): Teacher Symbol where
  membershipQuery := membershipQuery
  equivalenceQuery ldfa := (findCex membershipQuery nWords wordLen ldfa (mkStdGen (seed + ldfa.hash))).fst
  prop_L := h_target_regular
  prop_cex := by ...
\end{lstlisting}
\end{minipage}

The caller provides the membership query, i.e., the unknown language, together with a proof \lean{h_target_regular} that this language is regular.
As a membership query is an arbitrary Boolean function, its regularity cannot be derived from the teacher's implementation.
Section~\ref{sec:learn-dfa} shows how to obtain this proof for a concrete language.

For an equivalence query, the function \lean{findCex} generates \lean{nWords} random words of length \lean{wordLen}, sampling their symbols from the alphabet with the random generator.
While generating a word symbol by symbol, it runs the hypothesis alongside and compares its classification of every non-empty prefix with the membership query.
The first prefix that the two classify differently is returned as a counterexample.
If none of the \lean{nWords} words produces a counterexample, the equivalence query returns \lean{none}.

To ensure reproducibility while also varying the random words across learning rounds, the random generator is initialized with a fixed seed combined with a hash of the current hypothesis.
As this implementation serves as a proof of concept, the hashing function is kept simple and just hashes the number of states and the number of accepting states as shown in the Listing below.
Since the number of states grows in every round (Section~\ref{sec:termination}), the generated words usually differ between rounds.

\begin{lstlisting}
def LearnedDFA.hash (ldfa: LearnedDFA Symbol State): Nat :=
  Hashable.hash (ldfa.states.length, ldfa.accept.length) |>.toNat
\end{lstlisting}

The guarantee \lean{prop_cex} is proven for all inputs.
By induction over the length of the generated word, the state tracked by \lean{findCex} is always the state the hypothesis reaches with the current prefix.
Hence, a prefix is only returned as a counterexample if the hypothesis and the membership query disagree on it.

However, the random word teacher is not complete.
Since it only tests a finite number of words, it may miss a difference between the hypothesis and the unknown language and return \lean{none} for an incorrect hypothesis.
\lean{lstar} terminates with this teacher, and \lean{lstar_minimal} and \lean{lstar_unique} apply to its result, but \lean{lstar_correct} does not.
The result is only correct with respect to the performed tests.
A complete teacher requires an equivalence query with formal guarantees, such as the W-method of Chow~\cite{chow_testing_1978}, which we leave to future work.

\subsection{Learning a DFA from a Teacher}\label{sec:learn-dfa}

With \lean{createRandomWordTeacher} from Section~\ref{sec:random-word-teacher}, we can now run the Lean L*-algorithm implementation.
We use the alphabet \lean{Syms} from Section~\ref{sec:alphabet-repr} with the two symbols \lean{a} and \lean{b}.
The \lean{membershipQuery} in Line~\ref{ln:example-membershipQuery} of the Listing below defines the unknown regular language: all words with an even number of \lean{a}s and an even number of \lean{b}s, e.g., $\epsilon$, $aa$, $abba$.
It tracks the parities of both symbols with the transition function \lean{parityStep}.
As the language is given by a finite transition system, its regularity follows from the helper theorem \lean{isRegular_of_foldl} (Line~\ref{ln:example-regular}).
Finally, we create the teacher in Line~\ref{ln:example-teacher} with seed 0 and 500 random words of length 10 per equivalence query.

\noindent\begin{minipage}{\linewidth}
\begin{lstlisting}
def parityStep: Bool × Bool → Syms → Bool × Bool
  | (oddA, oddB), .a => (!oddA, oddB)
  | (oddA, oddB), .b => (oddA, !oddB)

def membershipQuery (w: List Syms): Bool :=(*@\label{ln:example-membershipQuery}@*)
  w.foldl parityStep (false, false) == (false, false)

theorem membershipQuery_isRegular: Language.IsRegular {w: List Syms | membershipQuery w} :=(*@\label{ln:example-regular}@*)
  isRegular_of_foldl parityStep (false, false) (· == (false, false))

def randomWordTeacher: Teacher Syms := createRandomWordTeacher membershipQuery membershipQuery_isRegular 0 500 10(*@\label{ln:example-teacher}@*)
\end{lstlisting}
\end{minipage}

To run this algorithm, we simply call:
\begin{lstlisting}
def dfaRandom: LearnedDFA Syms (List Bool) := lstar randomWordTeacher
\end{lstlisting}

This run finds the counterexamples $bb$ and $bab$ and ends with a 4-state hypothesis that is isomorphic to Figure~\ref{fig:learned-dfa-final}.
Since the random word teacher is not complete, \lean{lstar_correct} does not apply to \lean{dfaRandom}.
For this run, however, correctness follows after the fact: the result has 4 states, as many as the DFA given by \lean{parityStep} over \lean{Bool × Bool}, so \lean{lstar_unique} makes the two isomorphic, and the result accepts the target language.

\section{Conclusion}\label{sec:conclusion}

In this article, we present a mechanically verified, executable implementation of Angluin's L*-algorithm in Lean 4.
By constructing the algorithm together with its proofs of termination, minimality, and uniqueness for every teacher, and of correctness for every complete teacher, within a single, unified artifact, we demonstrate a methodology that closes the gap between formal specification and software.
The L* part of the Lean artifact spans around 2,400 lines of Lean code, excluding comments.
Even after considerable ``proof golfing'' to enhance clarity and theorem proving performance, the proofs still constitute roughly three-quarters of the code without comments.
The end-to-end theorems depend only on Lean's standard axioms (\lean{propext}, \lean{Classical.choice}, \lean{Quot.sound}), which our continuous integration checks together with the absence of \lean{sorry}.
The result is a single Lean artifact which simultaneously is an executable program and a proof.

\subsection{Findings from Formalization}\label{sec:findings}
Implementation and formalization yielded several key insights into the algorithm's structure and theoretical background.

\paragraph{Reordered Learning Round.}
A significant finding emerged from restructuring the learning round to simplify the termination proof.
In Angluin's presentation, a round first makes the observation table closed and consistent, then constructs the hypothesis, performs an equivalence query, and finally processes the counterexample.
We rotate this loop: every recursive call of \lean{lstarAux} starts with a closed and consistent observation table and receives a proof of these properties as an argument.
It then creates the hypothesis DFA, performs the equivalence query, processes the counterexample, and makes the updated observation table closed and consistent again before the next call.
The queries asked remain the same, but the rotation makes the invariant explicit, which immensely simplifies the formal argument.
The high-level termination argument relies on the hypothesis recognizing the counterexample being inequivalent to the old one.
For a formal proof, we need to construct this new hypothesis, which is only possible from a closed and consistent observation table.
Our reordered loop structure ensures we can do this in every learning round, proving progress towards an upper bound.
Moreover, the counterexample processing introduced by Angluin~\cite{angluin_learning_1987}, which adds the counterexample and all its prefixes to $S$, provably invalidates the consistency of the observation table.
Consequently, the subsequent call to \lean{makeClosedConsistent} is a provable starting point for strict progress.
We formalized this low-level argument as a separate theorem.
The termination proof of \lean{lstarAux} itself uses Angluin's high-level argument (Section~\ref{sec:termination}).

\paragraph{Upper Bound on the Number of States.}
Angluin takes the number $n$ of states of the minimum acceptor for the unknown regular language as given, which is standard in a pen-and-paper proof.
In a formalization, this bound has to be derived.
We based the termination proof, which requires the upper bound, on the Myhill-Nerode congruence (see Section~\ref{sec:myhill-nerode}), and showed in Section~\ref{sec:myhill-nerode-bound} how to obtain the upper bound from a given regular language.
It would simplify the termination proof if we designed the algorithm with this upper bound as its fuel, i.e., if the fuel decreased in every learning round, resulting in a trivial termination measure.
However, since \lean{maxStates} is \lean{noncomputable}, we cannot use it as fuel in executable code but only in the termination measure, which is only used for proving and not during compilation.

\paragraph{Computable Structures and mathlib.}
Using structures and lemmas from mathlib significantly simplified theorem proving, as we were able to reuse many of them in our code.
However, as mathlib is mostly designed with theorem proving in mind, many of its structures, such as sets, are not computable, and structures building on them, such as the accepting states of a mathlib \lean{DFA}, are not either.
We therefore had to create computable structures ourselves, such as \lean{LearnedDFA}, and specify a mapping to the mathlib structures for use in proofs.
In the end, this additional effort still pays off, given the large amount of actively maintained lemmas in mathlib.
To quantify this, we collected the mathlib declarations used by the proof terms of our L* implementation: it directly uses 175 declarations from 56 mathlib files, and its main theorems depend transitively on 2,349 mathlib declarations from 188 files.
Only 24 of them concern automata (DFAs, languages, and left quotients).
The directly used declarations mostly concern sets, finite sets, cardinalities, and lists; transitively, the general order and algebraic hierarchies of mathlib dominate.
Equally important, our theorem statements build on mathlib's standard definitions of DFAs, languages, and regularity; our own definitions occurring in them, such as \lean{LearnedDFA.accepts}, are thin wrappers around these, so the statements can be checked without trusting a custom automata theory.

\paragraph{Prefix- and Suffix-Closedness.}
During our work on the implementation, we also questioned whether the requirements that the set $S$ is prefix-closed and the set $E$ is suffix-closed can be dropped.
Avoiding those premises would help us reduce the number of membership queries, especially for the set $S$, as we would not need to add all prefixes during counterexample processing.
Leveraging the development environment of Lean, we could quickly find out where these premises are used in the formal proof: prefix-closedness only in the proof that every state of the hypothesis is reachable and in the proof of Lemma~\ref{thm:eval-eq-row}, suffix-closedness only in the proof of Lemma~\ref{thm:learned-dfa-consistent-with-T}.
Everywhere else, they only have to be preserved.
However, it turned out that both requirements are needed for the baseline algorithm and the counterexample processing introduced by Angluin:
\begin{itemize}
    \item Without prefix-closedness of $S$, a DFA created from a closed and consistent observation table can have unreachable states, and Lemma~\ref{thm:eval-eq-row} fails.
    For example, let $A = \{a, b\}$, the unknown language $\{ab\}$, $S = \{\epsilon, ab\}$, and $E = \{\epsilon\}$.
    The observation table is closed and consistent, but the state $row(ab)$ is unreachable: the hypothesis reaches $row(\epsilon)$ with $ab$ and rejects $ab$, although $T(ab) = 1$.
    \item Without suffix-closedness of $E$, two rows can be equal although their prefixes have different future behavior, such that the DFA created from the observation table is no longer consistent with it, i.e., Lemma~\ref{thm:learned-dfa-consistent-with-T} fails.
    For example, let $A = \{a\}$, the unknown language $\{w \mid |w| \geq 2\}$, $S = \{\epsilon\}$, and $E = \{\epsilon, aa\}$.
    The rows of $\epsilon$ and $a$ are both $(0, 1)$, so the observation table is closed and consistent, but the hypothesis has a single, rejecting state and rejects $aa$, although $T(aa) = 1$.
\end{itemize}

\paragraph{Use of LLMs.}
We also made use of LLMs.
During the original implementation, the models were not ready for whole proof generation: they hallucinated built-in and mathlib lemmas and produced \lean{sorry} when they could not complete a proof, so we only used manually checked, selected ideas.
In a later revision, we used an LLM-based coding agent (Claude Code with Claude Opus 5.5) to restructure the guarantees of the teacher and parts of the learner, to add the end-to-end theorems on correctness, minimality, and uniqueness of the learned DFA as well as the low-level termination argument, to prove the guarantees of the random word teacher, to write the examples, and to restructure and clean up the artifact repository.
The core of the formalization, i.e., the observation table, the learner, Angluin's lemmas and Theorem~\ref{thm:minimality}, and the termination proof of the algorithm, was written by the authors.
All proofs, including the generated ones, are checked by Lean's kernel, and the authors reviewed every change.
As motivated in the introduction, when parts of a program and its proofs are written by an AI, Lean guarantees that no change breaks a proven property.
Remaining for the human reviewer is checking that the theorem statements express the intended properties.
The coding agent also drafted and revised parts of the text of this article, in particular of Sections~\ref{sec:teacher-impl},~\ref{sec:lstar-proofs},~\ref{sec:run-lstar}, and~\ref{sec:conclusion}, and assisted in checking the related literature.
The authors reviewed and edited all text and take full responsibility for it.

\subsection{Related Work}
Our work covers the L*-algorithm introduced by Angluin~\cite{angluin_learning_1987}.
Many variants and extensions exist that build on this algorithm, e.g., improved counterexample processing~\cite{rivest_inference_1989}, learning of Mealy machines~\cite{shahbaz_inferring_2009} and Markov decision processes~\cite{tappler_l-based_2021}, and the L\# algorithm~\cite{vaandrager_new_2022}.
For all of those algorithms, correctness and termination were originally shown by pen-and-paper proofs.
Efficient implementations of those algorithms are provided by libraries like AALpy~\cite{muskardin_aalpy_2022} for Python and LearnLib~\cite{raffelt_learnlib_2005} for Java, but these implementations are not formally verified.
Recently, Wi{\ss}mann~\cite{wismann_formalized_2026} modeled learners and teachers coalgebraically in Agda and proved upper and lower bounds on the number of queries, among others for L\# on Mealy machines.
Termination follows from these bounds, and correctness is part of the game semantics: the learner has succeeded once the teacher accepts its hypothesis.
In contrast to our work, Wi{\ss}mann focuses on query complexity rather than on properties of the learned automaton, such as its minimality and uniqueness, and leaves L* for future work.
To the best of our knowledge, our work is the first mechanized proof of correctness and termination of the L*-algorithm.

Interactive theorem provers have been used for verifying large software systems, such as seL4~\cite{klein_sel4_2009}.
seL4 is an operating system kernel whose C implementation is verified in Isabelle/HOL against an abstract specification.
In this case, implementation and proofs are written separately and connected to each other afterwards.
Similarly, the KeY prover~\cite{ahrendt_deductive_2016} verifies existing Java code against its specification.
While attempting to verify the TimSort implementation of OpenJDK with KeY, de Gouw et al.~\cite{de_gouw_openjdks_2015} found a bug that crashes the standard library's sorting method, and proved a fixed version free of this bug.
Another approach is to do the formal development first and generate code from it.
Isabelle/HOL provides a code generator that translates functional specifications to languages like Haskell, OCaml, or Scala~\cite{haftmann_code_2010}.
Rocq (formerly Coq) supports the extraction of functional programs from its proofs~\cite{letouzey_new_2003}.
In both cases, the formal model and the executable program are written in different programming languages, and the translation between them has to be trusted.
Lean 4~\cite{moura_lean_2021} allows proving properties directly on the implementation, which is compiled to executable code by Lean's compiler.
We still have to trust the compiler, but we avoid the step of translating the formal model into a separate executable program in another programming language.
This approach is taken by the open-source project seLe4n~\cite{noauthor_sele4n_2026} for a microkernel inspired by seL4.
Here, the kernel transitions are implemented as executable Lean functions and their invariants are proven in Lean.

In our implementation we build on automata theory from mathlib~\cite{the_mathlib_community_lean_2020}, which provides DFAs, regular languages, and the Myhill-Nerode theorem~\cite{noauthor_mathlibcomputabilitymyhillnerode_2026}.
CSLib~\cite{barrett_cslib_2026} aims to provide a similar library for computer science, including formally verified code.
LeanMachines~\cite{carbonneau_leanmachines_2026} uses Lean 4 for state-based modeling with Event-B style refinement, where machines cannot be constructed without discharging their proof obligations.
We follow a similar principle.
Our \lean{LearnedDFA}, \lean{ObservationTable}, and \lean{Teacher} structures carry their properties as fields, such that each instance can only be constructed with proofs of its well-formedness.

\subsection{Future Work}
This work lays a solid foundation for several promising extensions to explore the trade-offs between proofs, generality, and performance:

\begin{itemize}
    \item \textbf{Extending the Learning Framework:}
    We can generalize our architecture to support a wider array of learning scenarios and automaton types.
    This includes implementing other learning algorithms like L\#~\cite{vaandrager_new_2022} or even passive learning algorithms like RPNI~\cite{oncina_inferring_1992}, as well as incorporating alternative counterexample processing strategies, such as the one proposed by Rivest and Schapire~\cite{rivest_inference_1989}.
    Additionally, we can expand the implementation to also learn different types of automata, like Mealy machines~\cite{shahbaz_inferring_2009}.
    
    \item \textbf{Verified Teacher Implementations:}
    Our random word teacher is sound but not complete, so \lean{lstar_correct} does not apply to it.
    A complete teacher requires an equivalence check with formal guarantees, like the W-method~\cite{chow_testing_1978}.
    Given an upper bound on the number of states of the target, the W-method is complete for all hypotheses up to that bound, which is exactly the form of completeness that \lean{Teacher.Complete} requires.
    Formally proving this in Lean would yield a fully self-contained, verified learning framework, which we could further extend to learn stochastic systems~\cite{tappler_l-based_2021}.
    
    \item \textbf{Performance Analysis and Optimization:}
    Given that Lean compiles to C and further to native machine code, a compelling research question is how the performance of this verified implementation compares to established, unverified libraries like AALpy~\cite{muskardin_aalpy_2022} or LearnLib~\cite{raffelt_learnlib_2005}.
    This also includes exploring the interplay between proof maintenance and performance optimization.
    How can a verified implementation be heavily optimized while preserving its formal proof structure?
    Answering this would provide valuable insights into the engineering of high-assurance, high-performance software.
\end{itemize}

\section*{Data Availability}
The Lean artifact (implementation, proofs, examples, and scripts reproducing the figures of this article) is archived on Zenodo~\cite{mascherbauer_aalean_2026}.
It builds with Lean \texttt{v4.36.0-rc2} and mathlib commit \texttt{0c07479f}.
Its README maps the sections, theorems, and lemmas of this article to the Lean declarations and lists, for each claim, the command that reproduces it.

\begin{acks}
Claude Code (with Claude Opus 5.5) was used to generate code and proofs of the artifact and to draft and revise parts of the text of this article (see Section~\ref{sec:findings}). 
All generated proofs were reviewed by the authors and checked by Lean.
All generated text was reviewed and edited by the authors.
\end{acks}

\bibliographystyle{ACM-Reference-Format}
\bibliography{references}

\end{document}